\documentclass[11pt]{article}
\usepackage{appb,epsfig}
\begin{document}
\def\lsim{\buildrel <\over\sim}
\def\gsim{\buildrel >\over\sim}
\eqsec  
\title{FERMI AND NON--FERMI LIQUID BEHAVIOR IN QUANTUM IMPURITY 
SYSTEMS:\\
CONSERVING SLAVE BOSON THEORY\thanks{Lectures presented at
the XXXVIII Cracow School of Theoretical Physics, June 1--10, 1998, 
Zakopane, Poland.}
}
\author{Johann Kroha and Peter W\"olfle
\address{Institut f\"ur Theorie der Kondensierten Materie,
Universit\"at Karlsruhe, Postfach 6980, 76128 Karlsruhe, Germany}
}
\maketitle
\begin{abstract}
The question of Fermi liquid vs. non--Fermi
liquid behavior induced by strong correlations is one of the
prominent problems in metallic local moment systems. As standard
models for such systems, the SU(N)$\times$SU(M) Anderson impurity
models exhibit both Fermi liquid and non--Fermi liquid behavior,
depending on their symmetry. Taking the Anderson model as an
example, these lectures first give an introduction to the auxiliary 
boson method to describe correlated systems governed by a strong,
short--range electronic repulsion. It is then shown how to include the 
relevant low--lying excitations (coherent spin flip and charge fluctuation
processes), while preserving the local gauge symmetry of the model.  
This amounts to a conserving T--matrix approximation (CTMA). 
We prove a cancellation theorem showing that the CTMA incorporates
all leading and subleading infrared singularities at {\em any}\/ given
order in a self--consistent loop expansion of the free energy.  
As a result, the CTMA recovers the correct infrared behavior of the 
auxiliary particle propagators, indicating that it correctly describes 
both the Fermi and the non--Fermi regimes of the Anderson model.
\end{abstract}
  
\section{Introduction}
It is a remarkable feature of interacting, itinerant 
fermion systems that at low temperatures $T$ they behave in 
general in much the same way as a noninteracting Fermi gas,
even though the interaction may be strong. An extremely successful
description of this phenomenon, known as Fermi liquid (FL) behavior, 
is provided by the notion of quasiparticles, which was established
by Landau's phenomenological Fermi liquid theory \cite{landau.53}. 
The key assumption is that, as the interaction is continuously
turned on, there exists a 1:1 correspondence between the low energy 
eigenstates of the interacting system and the single--particle 
states of the free Fermi gas. Therefore, the low--lying interacting 
states may be described approximately as single--particle states or 
quasiparticles, whose decay rate $1/\tau$ is small compared to their 
excitation energy $\omega$, $1/\tau \ll \omega$, and which are 
characterized by the same quantum numbers as the noninteracting 
states. As a consequence, Fermi liquids exhibit the
same low--$T$ thermodynamics as a noninteracting Fermi system,
e.g.~a linear in $T$ specific heat $c=\gamma T$ and a
constant Pauli paramagnetic susceptibility $\chi _o$. However,
the effective mass and other parameters may be renormalized by
the interaction, resulting in an enhancement of the 
specific heat coefficient $\gamma$ and the susceptibility 
$\chi _o$. It is at the heart of the quasiparticle picture
that at low $T$ the Pauli exclusion principle substantially
reduces the phase space available to quasiparticle scattering.
This blocking mechanism is effective as long as the 
quasiparticle interaction is {\it shortranged in space and time}, 
which is usually the case in three dimensions because of screening.
It also implies that the quasiparticle scattering rate
vanishes as $1/\tau \propto (\omega ^2 + T^2)$ in the limit 
$\omega,\; T\to 0$, thus providing
a microscopic justification for the basic assumption of FL theory
and leading to an interaction contribution to the electrical
resistivity which behaves as $\Delta \rho \propto T^2$. 
Obviously, the Pauli principle as the origin of FL behavior is
very robust, which explains the almost ubiquitous presence of
a FL ground state in interacting Fermi systems and the broad
success of Fermi liquid theory.

In this light it is all the more exciting that in recent years
a number of new alloys have been discovered which exhibit 
striking deviations from this usual behavior. 
These systems have in common that a localized,
degenerate degree of freedom, the magnetic moment of a magnetic 
ion, is dynamically coupled to a continuum of conduction
electron states. In general, such a coupling generates the Kondo
effect, characterized by resonant spin flip scattering of electrons 
at the Fermi surface off the local moment. Concomitantly, the
conduction electron spin flip rate initially increases 
logarithmically as the temperature is lowered, passes 
through a maximum at a characteristic scale, the Kondo temperature 
$T_K$, and approaches zero as $T\rightarrow 0$, because the 
effective local moment becomes screened by the conduction electron 
spins. Thus, even for many strongly correlated systems of this type a 
Fermi liquid description applies below $T_K$, with usually a strongly 
enhanced quasiparticle effective mass, lending the term
``heavy fermion systems'' to these materials. 

Completely new physics may arise, however, if the quenching of 
the local moments is inhibited. Two different mechanisms for the 
appearance of a non--Fermi liquid ground state have been put forward: 
\begin{itemize}
\item[(1)] Proximity of a quantum phase transition (QPT) to an
antiferromagnetically ordered state \cite{millis.93}--\cite{sachdev.95}
as a function of a dopant concentration $x$ or of pressure.
Near the QPT the quantum critical fluctuations become 
longranged in space and time and can, thus, mediate a longrange
quasiparticle interaction, leading to a breakdown of FL
theory. There are indications for this spatially extended 
mechanism to be realized near the QPT of certain Ce based compounds 
like CeCu$_{6-x}$Au$_x$ \cite{loehn.94,loehn.96}, 
CeCu$_2$Si$_2$ \cite{steglich.96}, and CePd$_2$Si$_2$ 
\cite{lonzar.96a,lonzar.96b}. 
\item[(2)] Two--channel Kondo effect (2CK) \cite{nozieres.80,coxzawa.98}. 
The local magnetic moment is coupled to two exactly degenerate conduction
electron channels. Because of a frustration effect between 
the screening of the local moment by the different conduction
channels the moment quenching cannot be complete, leading to 
a nonvanishing conduction electron spin scattering rate even 
at the lowest temperatures and subsequently to a breakdown of FL behavior. 
It has been suggested \cite{cox.87} that this mechanism, 
based on single--ion dynamics rather than longrange fluctuations, may be
realized predominantly in U based materials with cubic 
symmetry about the magnetic ion, such as Y$_{1-x}$U$_x$Pd$_3$
\cite{maple.94} or UCu$_{5-x}$Pt$_x$ \cite{maple.96}, 
which do not exhibit a QPT.
\end{itemize}
\noindent

In both scenarios the wealth of experimental data
showing non--FL behavior at low temperatures is not consistently
explained by the present theories. Open questions within the
QPT picture include, e.g., whether the local impurity
dynamics competing with the magnetic ordering can play a role,
and how the transition from the spin screened heavy FL phase to the
magnetically ordered phase occurs alltogether.
In the 2CK mechanism, on the other hand, inter--impurity 
interactions could modify the single ion behavior. 
Exact solution methods as well as numerical simulations have 
provided important progress in our understanding of
strongly correlated quantum impurity systems. 
However, their applicability is essentially restricted 
to problems involving only a single impurity, owing to integrability 
conditions or limitations in the numerical effort, respectively. 
Therefore, more generally applicable theoretical techniques are 
called for.

In the present work we focus on the single--ion dynamics. 
We develop a standard field theoretical method,
based on an auxiliary particle or slave boson representation, 
which describes the quantum impurity dynamics in a controlled way and 
at the same time has the potential of being extended to problems of many
impurities on a lattice. As a standard model of strongly correlated 
electrons which, depending on its symmetry, exhibits both FL and 
non--FL behavior, we consider the SU(N)$\times$SU(M) Anderson impurity 
model of a local, $N$--fold degenerate degree of freedom, coupled 
to $M$ identical conduction bands.

In order to set the stage for the more formal development of the
theory, in the following section we will briefly review the
striking differences in the phenomenology of the single--channel
and the multi--channel Kondo effects. In section 3 the slave boson 
representation is introduced, which provides a particularly 
compact formulation of the SU(N)$\times$SU(M) Anderson model.
We also discuss why the presence of FL or non--FL behavior in a
given quantum impurity system can already be seen from the singular
infrared dynamics of the auxiliary particles. Section 4 contains a 
critical assessment of earlier approximate slave boson treatments. 
This will motivate our conserving slave boson approach (conserving
T--matrix approximation, CTMA), which is developed in section 5.
As will be seen, the results produced from this theory are in very 
good agreement with known exact properties of the model. 
Conclusions are drawn in section 6. In the appendices we prove a 
cancellation theorem for non--CTMA diagrams, which justifies the
CTMA on formal grounds, and derive in detail the self--consistent
CTMA equations.

\section{Single-- and multi--channel Kondo effect and possible
physical realizations}
\begin{figure}
\vspace*{-0cm}
\centerline{\psfig{figure=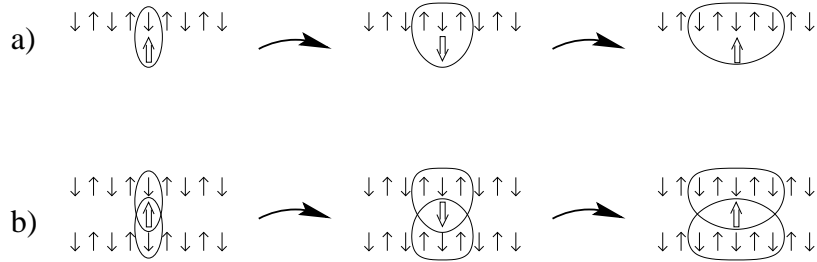,width=10.7cm}}
\vspace*{0.6cm}
\caption{
Sketch of the renormalization group for a) the single--channel Kondo 
model
(local moment compensation) and 
b) the two--channel Kondo model (local moment over--compensation). 
Small arrows  denote conduction electron spins 1/2, a heavy
arrow a localized spin 1/2. The curved arrows indicate successive 
renormalization steps.
\label{cartoon}}
\end{figure}

In this section we briefly discuss how the single--channel and the 
two--channel Kondo effects may arise in magnetic metals, if the 
interaction between the local moments can be neglected. We first 
discuss the usual magnetic, single--channel Kondo effect.  
A local moment is generated by an atomic f or d level
whose energy $E_d$ lies far below the Fermi energy $\varepsilon _F 
\equiv 0$ and whose electron occupation number is effectively 
restricted to $n_d  \le 1$ by a strong Coulomb repulsion $U$ 
between two electrons in the same orbital. While the angular 
momentum degeneracy of the level is usually lifted by crystal field
splitting, in the absence of a magnetic field a twofold degeneracy 
of a level occupied by one electron is guaranteed by time reversal 
symmetry (Kramers doublet), corresponding to the spin quantum 
numbers $m=\pm 1/2 $ of the electron. In addition, there is a 
hybridization matrix element $V$ between the atomic orbital and the
conduction electron states. Such a system is described by the
single impurity Anderson Hamiltonian
\begin{equation}
H=\sum _{\vec k,\sigma }\varepsilon _{\vec k}
c_{\vec k\sigma}^{\dag}c_{\vec k\sigma}+
E_d\sum _{\sigma} d_{\sigma}^{\dag}d_{\sigma}+
V\sum _{\vec k,\sigma }(c_{\vec k\sigma}^{\dag}d_{\sigma} +h.c.) + 
U d^{\dag}_{\uparrow}d^{\phantom{\dag}}_{\uparrow}
d^{\dag}_{\downarrow}d^{\phantom{\dag}}_{\downarrow} ,
\label{Ahamilton}
\end{equation}
where $c^{\dag}_{\vec k \sigma}$ and $d^{\dag}_{\sigma}$ are the 
creation operators of a conduction electron with dispersion 
$\varepsilon _{\vec k}$ and of an electron in the local orbital with spin
$\sigma $, respectively. The low energy physics of this system is 
dominated by processes of second order in $V$, by which an electron 
hybridizes with the conduction band and the impurity level is 
subsequently filled by another electron, thereby effectively flipping 
the impurity spin. Thus, in the region of low excitation energies, 
the Anderson Hamiltonian (\ref{Ahamilton}) may be mapped  
onto the s--d exchange (or Kondo) model \cite{schriwo.66}, 
the effective coupling between the impurity spin and the 
conduction electron spin always being antiferromagnetic: 
$J=|V|^2/|E_d| > 0$ ($U\gg |E_d|$). 
These models have been studied extensively by means
of Wilson's renormalization group \cite{wilson.75}, by the Bethe 
ansatz method \cite{andrei.83,wiegmann.83} and by means of a 
phenomenological Fermi liquid theory \cite{nozieres.74}. 
In this way the following physical 
picture has emerged: The model contains a dynamically generated low 
temperature scale, the Kondo temperature, which is expressed in terms 
of the parameters of the Anderson Hamiltonian (\ref{Ahamilton}) as
$T_K = D (N \Gamma / D )^{(M/N)}{\rm exp}\{ -\pi E_d/(N \Gamma)\}$, 
with ${\cal N}(0)$ and $D=1/{\cal N}(0)$ the density of states at the 
Fermi energy and the high energy band cutoff, respectively.
$\Gamma = \pi V^2 {\cal N}(0)$ denotes the effective hybridization
or $d$--level broadening and $N$, $M$ are the degeneracy
of the local level and the number of conduction electron channels 
(see below). In the intermediate temperature regime, $T \gsim T_K$,
resonant spin flip scattering of electrons at the Fermi surface 
off the local degenerate level leads to logarithmic contributions to the 
magnetic susceptibility, the linear specific heat coefficient and 
the resistivity ($\rho(T)=\rho(0)+\Delta \rho(T)$), 
$\chi(T),\ \gamma (T),\ \Delta\rho (T) 
\propto -{\rm ln}(T/T_K)$, and to a breakdown of perturbation theory 
at $T\simeq T_K$. Below $T_K$ a collective many--body spin singlet 
state develops in which the impurity spin is screened by the conduction 
electron spins as lower and lower energy scales are successively 
approached, leaving the system with a pure potential scattering center. 
The spin singlet formation is sketched in Fig.~\ref{cartoon} a) 
and corresponds to a vanishing entropy at $T=0$, $S(0)=0$. 
It also leads to saturated behavior of physical quantities 
below $T_K$, like $\chi (T)=const$, $c (T)/T =const.$ 
and $\Delta \rho (T)\propto T^2$, i.e. to Fermi liquid behavior.

As an example of possible two--channel Kondo systems we discuss  
the uranium based compounds mentioned in the introduction.
The U$^{4+}$ ions have nominally a 5f$^2$ configuration, i.e. an 
even number of electrons, which does not allow for a Kramers degenerate 
ground state because of integer total spin.
However, in the cubic crystal symmetry of these materials 
the orbital degeneracy may be not completely lifted, so that there can 
be an approximate twofold degeneracy of the U$^{4+}$ ground state, 
corresponding to two different orientations of the electrical quadrupole 
moment of the 5f orbital in the lattice (quadrupolar Kondo effect) 
\cite{coxzawa.98,cox.87}. This degree of freedom may be flipped by 
scattering of conduction electrons (which in the cubic symmetry also 
have a twofold angular momentum degeneracy). The conduction electron 
spin is conserved in this scattering process, leaving
it as a Kramers degenerate scattering channel degree of freedom, which 
we will label by $\mu = 1,...,M$, $M=2$. Describing the
orbital degree of freedom as a pseudospin 1/2, labelled by the quantum 
number $\sigma = 1,\dots ,N$, $N=2$, in analogy to the magnetic Kondo 
effect, we arrive at the SU(2)$\times$SU(M) symmetric Kondo model, 
\begin{equation}
H=\sum _{\vec k,\sigma,\mu}\varepsilon _{\vec k}
c_{\vec k\mu\sigma}^{\dag}c_{\vec k\mu\sigma}+
J \sum _{\vec k, \vec k ',\sigma,\sigma ',\mu}c^{\dag}_{\vec k \mu \sigma }
\vec S\cdot \vec \tau _{\sigma\sigma '}
c^{\phantom{\dag}}_{\vec k \mu \sigma '},
\label{2CKhamilton}
\end{equation}
where $\vec S$ is the local pseudospin operator and 
$\vec \tau _{\sigma\sigma'}$ the vector of Pauli matrices. 
To keep the naming uniform, we will refer to the orbital degree of 
freedom as the (pseudo)spin or local moment, $\sigma$, in analogy to 
the magnetic Kondo effect, and to the physical electron spin as the 
channel degree of freedom, $\mu$. In the multi--channel case, too, 
the conduction electrons of each channel {\it separately} screen the 
impurity moment by multiple spin scattering at temperatures below the 
Kondo scale $T_K$. However, in this case, the local moment is 
over--compensated, since the impurity spin can never form a singlet 
state with both conduction electron channels at the same time in this 
way, as can be seen in Fig.~\ref{cartoon} b). As a consequence of 
this frustration, there is not a unique ground state, leading to a 
finite residual entropy \cite{andrei.84,tsvelik.85} at $T=0$ of 
$S(0)=k_B {\rm ln}\sqrt{2}$ in the two--channel model. In particular, 
the precondition of FL theory of a 1:1 correspondence 
between interacting and non--interacting states is violated. 
As a consequence, characteristic singular temperature dependence 
\cite{andrei.84,affleck.91} of physical quantities persists 
for $T\lsim T_K$ down to $T=0$: $\chi (T)\propto -{\rm ln}(T/T_K)$, 
$c (T)/T  \propto -{\rm ln}(T/T_K)$ and 
$\rho (T)-\rho(0)\propto -\sqrt {T/T_K}$. Note, however, that this
behavior may be changed by any crystal field splitting of the 
quadrupolar non--Kramers ground state doublet.

In order to apply standard field theoretical methods to the 
multi--channel Kondo model it is convenient to consider it as the 
low--energy limiting case of a corresponding Anderson model as discussed 
for the single--channel case. Here, in addition, 
the conservation of the channel degree of freedom has to be guaranteed. 
This can be implemented in an elegant way using an auxiliary boson
representation to be discussed in the next section.

\section{Auxiliary particle representation} 
As discussed above, the local level of a quantum 
impurity in the limit of infinitely strong local Coulomb repulsion $U$ 
between electrons in the same level allows only for at most single electron
occupation of the level, $n_d\le 1$. One should note that for a realistic
finite value of $U$ the low--energy physics of the model is effectively
still confined to the part of the Hilbert space without multiple 
occupancy. Therefore, the model Eq.~(\ref{Ahamilton}) in the limit 
$U\rightarrow \infty$ is the generic model for the physics of quantum 
impurities at large $U$ in general. 

A powerful technique for implementing the projection in Hilbert space 
caused by a large Coulomb repulsion $U$ is the method of auxiliary 
particles (slave bosons, pseudofermions) \cite{barnes.76}. 
Each Fock state $|\alpha\rangle$ of the
impurity is assigned a creation operator, which can be envisaged
as creating the state out of a vacuum state $|{\rm vac}\rangle$ without
any impurity level at all, 
$|\alpha\rangle = a_{\alpha}^{\dag} |{\rm vac}\rangle$.
(E.g., for a single orbital there are four such states, 
$|0\rangle$ (empty orbital), $|\uparrow\rangle$ or $|\downarrow\rangle$
(orbital occupied by a single electron with spin $\uparrow$ or 
$\downarrow$) and $|2\rangle$ (level occupied by two electrons with 
spin $\uparrow$ and $\downarrow$).) Due to the requirements of Fermi 
statistics, the creation operators $a_{\alpha}^{\dag}$ are 
Fermi (Bose) operators for the states holding an odd (even) number of
electrons (or vice versa). The physical state corresponds to the sector 
of Fock space with exactly one auxiliary particle, 
$\sum _{\alpha}n_{\alpha} =1$, where 
$n_{\alpha}=a_{\alpha}^{\dag}a_{\alpha}$ is the occupation number
operator of particles $\alpha$.
Compared to alternative ways of effecting the projection, the auxiliary 
particle method has the advantage of making available the powerful
machinery of quantum field theory, provided the constraint on the total
auxiliary particle number can be incorporated in a satisfactory way.
 
For the quantum impurity models of the Anderson type introduced in the 
preceding section, only particles creating empty and singly occupied 
states are needed. We define $N$ pseudofermion creation operators 
$f_{\sigma}^{\dag}$ for each of the singly occupied states (labelled 
by $\sigma =1,2,\dots , N$) and $M$ boson creation operators 
$b_{\bar\mu}^{\dag}$ for each of the empty states created when an electron 
hops from the impurity into the $\mu$-th conduction electron band 
(labelled by $\mu = 1,2,\dots , M$). In terms of these
operators the Hamiltonian of the SU(N)$\times$SU(M) Anderson model 
Eq.~(\ref{Ahamilton}) takes the form
\begin{equation}
H=\sum _{\vec k,\sigma ,\mu}\varepsilon _{\vec k}
c_{\vec k\mu\sigma}^{\dag}c_{\vec k\mu\sigma}+
E_d\sum _{\sigma} f_{\sigma}^{\dag}f_{\sigma}+
V\sum _{\vec k,\sigma ,\mu}(c_{\vec 
k\mu\sigma}^{\dag}b_{\bar\mu}^{\dag}f_{\sigma} +h.c.)
\label{sbhamilton}
\end{equation}
In order for $H$ to be SU(M) invariant, the
slave boson multiplet $b_{\bar\mu}$ transforms according to the conjugate 
representation of the SU(M). In addition, the operator constraint
\begin{equation}
Q\equiv \sum _{\sigma} f_{\sigma}^{\dag}f_{\sigma} +
        \sum _{\mu} b_{\bar\mu}^{\dag}b_{\bar\mu} =1
\end{equation}
has to be satisfied at all times. One might interpret the constraint as
a statement of charge quantization, with the integer $Q$ the conserved,
quantized charge. Similar to quantum field theories with conserved 
charges, the charge conservation is intimately related to the existence 
of a local gauge symmetry. Indeed, the system defined by the Hamiltonian 
Eq.~(\ref{sbhamilton}) is invariant under simultaneous local $U(1)$ 
gauge transformations $f_{\sigma}\rightarrow f_{\sigma} {\rm e}^{i\phi 
(\tau )}$, $b_{\bar\mu}\rightarrow b_{\bar\mu} {\rm e}^{i\phi (\tau )}$, with 
$\phi (\tau )$ an arbitrary time dependent phase. 

\subsection{Exact projection onto the physical Hilbert space}
While the gauge symmetry guarantees the conservation of the quantized 
charge $Q$, it does not single out any particular $Q$, such as $Q=1$. 
In order to effect the projection onto the sector of Fock space 
with $Q=1$, one may use a procedure first proposed by
Abrikosov \cite{abrikosov.65}: Consider first the grand--canonical 
ensemble with respect to $Q$, defined by the statistical operator
\begin{equation}
\hat \rho _G = \frac{1}{Z_G} {\rm e}^{-\beta (H+\lambda Q)},
\label{eq:stat_op}
\end{equation}
where $Z_G={\rm tr}[{\rm exp}\{-\beta (H+\lambda Q)\}]$ 
is the grand--canonical partition function with respect to
$Q$, $-\lambda$ is the associated chemical potential,
and the trace extends over the complete Fock space,
including summation over $Q$. The expectation value of an observable
$\hat A$ in the grand--canonical ensemble is given by
\begin{equation}
\langle \hat A\rangle _G = {\rm tr} [\hat \rho _G \hat A] .
\label{eq:exp_val}
\end{equation}
The physical expectation value of $\hat A$, $\langle \hat A\rangle$, 
is to be evaluated in the canonical ensemble where $Q=1$. It can be 
calculated from the grand--canonical ensemble by differentiating with 
respect to the fugacity $\zeta = {\rm e}^{-\beta\lambda}$ and
taking $\lambda$ to infinity \cite{coleman.84},
\begin{equation}
\langle \hat A\rangle =
\lim _{\lambda \rightarrow \infty}
\frac {\frac{\partial }{\partial \zeta} \mbox{tr}
       \bigl[\hat A e^{-\beta (H+\lambda Q)} \bigr]} 
      {\frac{\partial }{\partial \zeta} \mbox{tr}
       \bigl[ e^{-\beta (H+\lambda Q)} \bigr]} =
\lim _{\lambda \rightarrow\infty}\frac{\langle Q\hat A\rangle _G}
{\langle Q \rangle _G}\ .
\label{projection}  
\end{equation}
{\em Projecting operators acting on the impurity states:}\/--- 
We list two important results, which follow straightforwardly
from Eq. (\ref{projection}): First,
the canonical partition function in the subspace $Q=1$ is
\begin{eqnarray}
Z_C & = & \lim _{\lambda \rightarrow \infty} \mbox{tr}
          \bigl[ Q e^{-\beta (H+\lambda (Q-1))} \bigr] \nonumber \\ 
    & = & \lim _{\lambda \rightarrow \infty}
          \bigl( e^{\beta\lambda} \langle Q \rangle _{G}(\lambda ) 
          \bigr)Z_{Q=0}\ \ \ ,\label{eq:zcan}
\end{eqnarray}
where the subscripts $G$ and $C$ denote the grand--canonical and the
canonical ($Q=1$) expectation value, respectively. Second, the 
canonical $Q=1$ expectation
value of any operator $\hat A$ which has a zero expectation value in 
the $Q=0$ subspace, $\hat A|Q=0\rangle =0$, is given by,
\begin{equation}
\langle \hat A \rangle _C = \lim _{\lambda \rightarrow \infty}
\frac {\langle \hat A \rangle _{G}(\lambda )}{ 
\langle Q \rangle _{G}(\lambda )}
\label{eq:canonical-expectation}
\end{equation}
Note that $\hat A|Q=0\rangle =0$ holds true for any physically observable
operator acting on the impurity. Examples are the physical electron operator 
$d^{\dag}_{\mu\sigma}=f^{\dag}_{\sigma}b_{\bar\mu}$ or the local spin
operator 
$\vec S = \sum _{\sigma\sigma '}
\frac{1}{2}f^{\dag}_{\sigma}\vec \tau _{\sigma\sigma '}f_{\sigma '}$.
In this case the operator $Q$ appearing in the numerator  
of Eq. (\ref{projection}) is not necessary to project away the $Q=0$ 
sector. In particular, the constrained $d$--electron Green's function is 
given in terms of the grand--canonical one ($G_{d}(\omega ,T,\lambda )$) 
as
\begin{equation} 
G_d(\omega ) = \lim _{\lambda \rightarrow \infty}
\frac {G_{d}(\omega ,T,\lambda )}{\langle Q \rangle _{G} (\lambda )} 
\label{eq:GdNCA}
\end{equation}
In the enlarged Hilbert space ($Q=0,1,2,...$) $G_{d}(\omega, T, \lambda )$ 
may be expressed in terms of the grand--canonical
pseudo--fermion and slave boson Green's functions 
using Wick's theorem. These auxiliary particle Green's functions, which
constitute the basic building blocks of the theory, are defined  
in imaginary time representation as
\begin{eqlettarray}
{\cal G}_{f\sigma}(\tau_1-\tau_2) & = - \langle T \{f_\sigma
(\tau_1)f_\sigma^{\dag}(\tau_2)\}\rangle_G \\
{\cal G}_{b\bar\mu}(\tau_1 - \tau_2) & = - \langle T\{ b_{\bar\mu} (\tau_1)
b_{\bar\mu}^{\dag}(\tau_2)\}\rangle_G\ ,
\eqalabel{eq:def_gfb}
\end{eqlettarray}
where $T$ is the time ordering operator.  The Fourier transforms of
${\cal G}_{f,b}$ may be expressed in terms of the exact self--energies
$\Sigma_{f,b}$ as
\begin{equation}
{\cal G}_{f,b}(i\omega_n) = \Big\{[{\cal G}_{f,b}^0(i\omega_n)]^{-1} -
\Sigma_{f,b}(i\omega_n)\Big\}^{-1}
\label{green}
\end{equation}
where
\begin{eqlettarray}
{\cal G}_{f\sigma}^0(i\omega_n) &=& (i\omega_n - E_d - \lambda)^{-1}\\
{\cal G}_{b\bar\mu}^0(i\omega_n) &=& (i\omega_n - \lambda)^{-1} 
\eqalabel{green0}
\end{eqlettarray}
Since as a consequence of the projection 
procedure $\lambda \rightarrow \infty$ the energy eigenvalues 
of $H + \lambda Q$  scale to infinity as $\lambda Q$, 
it is useful to shift the zero of the auxiliary particle
frequency scale by $\lambda$ (in the $Q = 1$ sector) and to define 
the ``projected'' Green's functions as 
\begin{equation}
G_{f,b}(\omega ) = \lim_{\lambda\to\infty} {\cal G}_{f,b}(\omega +
\lambda)
\label{eq:def_gfbproj}
\end{equation}
Note that this does not affect the energy 
scale of physical quantities (like the local $d$ electron Green's 
function), which is the {\it difference} between the 
the pseudo--fermion and the slave--boson energy.

{\em Canonical expectation values of conduction electron 
operators:}\/--- 
The canonical (i.e. projected onto the $Q=1$ subspace),
local conduction electron Green function is given as
\begin{equation}
G_{c\mu\sigma}(i\omega_n) = \Big\{[G_{c\mu\sigma}^0(i\omega_n)]^{-1} -
\Sigma_{c\mu\sigma}(i\omega_n)\Big\}^{-1}
\label{eq:greenc}
\end{equation}
with 
\begin{equation}
G_{c\mu\sigma}^0 (i\omega_n) = \sum _{\vec k} G^0_{c\mu\sigma}
(\vec k,i\omega_n)=
\sum _{\vec k} (i\omega_n +\mu _c - \epsilon_{\vec k}) ^{-1}\ ,
\label{eq:greenc0}
\end{equation}
where $\mu _c$ is the chemical potential of the conduction electrons.
The canonical, local $c$--electron self--energy,
$\Sigma_{c\mu\sigma} (i\omega _n)$, cannot be obtained from the
grand--canonical one by simply taking the limit $\lambda \rightarrow 
\infty$, since the $c$-electron density has a non--vanishing expectation 
value in the $Q=0$ subspace. However, it follows immediately from
the Anderson Hamiltonian (\ref{Ahamilton}), that the exact, canonical
conduction electron t--matrix $t_{\sigma\mu}(i\omega )$, defined by
$G_c = G^0_c [ 1+tG^0_c ]$, is   
is proportional to the full, projected $d$--electron Green's function,
$t_{\mu\sigma}(i\omega ) = |V|^2G_{d\mu\sigma}$.
Thus, we have as an exact relation,
\begin{equation}
G_{c\mu\sigma} (i\omega_n) =
G_{c\mu\sigma}^0 (i\omega_n) \Big[ 1 +  |V|^2 G_{d\mu\sigma} 
(i\omega_n) G_{c\mu\sigma}^0 (i\omega_n) \Big] \  ,
\end{equation}
and by comparison with Eq. (\ref{eq:greenc}) 
we obtain the local conduction electron self--energy respecting the
constrained dynamics in the impurity orbital,
\begin{equation}
\Sigma_{c\mu\sigma} (i\omega_n) = \frac {|V|^2 G_{d\mu\sigma} 
(i\omega_n)}
{1+|V|^2 G_{c\mu\sigma}^0 (i\omega_n) G_{d\mu\sigma} 
(i\omega_n)}\ . 
\label{sigmac}
\end{equation}
Using phenomenological Fermi liquid theory \cite{nozieres.74} and 
also by means of perturbation theory to infinite order in 
the on--site repulsion $U$ \cite{yamada.75} it has been shown for 
the Fermi liquid case $M=1$ of the symmetric Anderson model 
($2E_d = -U$) that the exact $d$--electron
propagator $G_{d\sigma}(\omega )$ and the $d$--electron self--energy
$\Sigma _{d\sigma}(\omega )\equiv \omega - G_{d\sigma}(\omega )^{-1}$ 
obey the following local Fermi liquid relations 
in the limit $\omega \to 0-i0$, $T \to 0$,
\begin{eqnarray}
\mbox{Luttinger theorem:\ \ }&&
\int  d\omega\; f(\omega )
\frac{\partial \Sigma _{d\sigma}(\omega )}{\partial \omega}\;
G_{d\sigma }(\omega )=0
\label{luttinger}\\
\mbox{Friedel--Langreth:\ \ }&&
\frac{1}{\pi}{\rm Im} G_{d\sigma }(\omega ) = \frac{1}{\Gamma} \sin ^2
\Big(\frac{\pi n_d}{N}\Big) - c\; 
\Big[\Big(\frac{\omega}{T_K}\Big)^2+\Big(\frac{\pi T}{T_K}\Big)^2\Big]
\nonumber\\
\label{friedel}\\
&&\hspace*{-3cm}
{\rm Im} \Sigma _{d\sigma}(\omega ) = 
\frac{\Gamma}{\sin ^2(\pi n_d/N)} +
c \Big(\frac{\Gamma}{\sin ^2(\pi n_d/N)}\Big)^2  
\Big[\Big(\frac{\omega}{T_K}\Big)^2+\Big(\frac{\pi T}{T_K}\Big)^2\Big],
\nonumber\\
\label{friedel_sigma}
\end{eqnarray}
where $c$ is a constant of $O(1)$.
Combining Eqs.~(\ref{sigmac}), (\ref{friedel}) it follows that 
(for $M=1$, away from particle hole symmetry) 
$\Sigma _{c\sigma}$ exhibits (in an exact theory) 
local Fermi liquid behavior as well, 
$\mbox{Im}\Sigma_{c\sigma}(\omega -i0,T=0) =a + b (\omega /T_K)^2$ 
for $\omega\rightarrow 0$. Note that this quantity is different from
the grand--canonical conduction electron self--energy and has a 
finite imaginary part at the Fermi level.

The momentum dependent conduction electron Green's function in the 
presence of a single impurity is given in terms of the canonical 
$d$--electron propagator as
\begin{equation}
G_{c\mu\sigma}(\vec k,\vec k{'}\;;\;i\omega_n) = G_{c\mu\sigma}^0
(\vec k,i\omega_n)
\Big[\delta_{\vec k,\vec k{'}} + |V|^2 G_d(i\omega_n)
G_{c\mu\sigma}^0(\vec k{'},i\omega_n)\Big]\ .
\end{equation}
The latter expression is the starting point for treating a random
system of many Anderson impurities \cite{coxwilkins.87}.
\\

\subsection{Analytical properties and infrared behavior}
The Green's functions $G_{f,b,c}$ have the following spectral 
representations
\begin{equation}
G_{f,b,c}(i\omega_n) = \int_{-\infty}^\infty d\omega{'}
\frac{A_{f,b,c}(\omega{'})}{i\omega_n - \omega{'}}
\end{equation}
with the normalization of the spectral functions $A_{f,b,c}$
\begin{equation}
\int_{-\infty}^\infty d\omega A_{f,b,c}(\omega) = 1.
\end{equation}
Taking the limit $\lambda \rightarrow \infty$ has important 
consequences on the analytical structure of the auxiliary particle 
Green's functions:

(1) ---
It follows directly from the definitions Eqs. (\ref{eq:def_gfbproj}),
(\ref{eq:def_gfb}), using Eqs. (\ref{eq:stat_op}), (\ref{eq:exp_val}),
that the traces appearing in the canonical functions $G_{f,b}$ 
are taken purely over the $Q=0$ sector of Fock space 
\footnote{This means that the auxiliary particle propagators 
are {\em not}\/ calculated in the canonical ($Q=1$) ensemble. 
The projection onto the $Q=1$ sector of Fock space is achieved only 
when they are combined to calculate expectation values of physically 
observable operators like $G_{d\sigma}$, $\langle\vec S\rangle$ etc.
The latter can be seen explicitly, e.g., from Eq.~(\ref{gdNCA}), 2nd equality.}.
Thus, the backward--in--time ($\tau _1 < \tau _2$) or hole--like 
contribution to the auxiliary propagators in 
Eq. (\ref{eq:def_gfb}) vanishes after projection, and we have
\begin{eqlettarray}
G_{f\sigma}(\tau_1-\tau_2) &=& -\Theta (\tau_1-\tau_2)
\lim _{\lambda\to \infty}
\langle f_{\sigma} (\tau_1)f^{\dag}_{\sigma} (\tau_2)\rangle _G\ \\
G_{b\bar\mu}(\tau_1-\tau_2) &=& -\Theta (\tau_1-\tau_2)
\lim _{\lambda\to \infty}
\langle b_{\bar\mu} (\tau_1)b^{\dag}_{\bar\mu} (\tau_2)\rangle _G\ .
\eqalabel{eq:def_gfb1}
\end{eqlettarray}
Consequently, their spectral functions $A_{f,b}$ have the Lehmann 
representation
\begin{eqlettarray}
A_{f\sigma} (\omega )& = &\sum _{m,n\geq 0} \mbox{e}^{-\beta E_m^0}
\mid \langle 1,n | f_{\sigma}^{\dag} | 0,m\rangle \mid^2
\delta (\omega -( E_n^1 - E_m^0 )) 
\label{lehmannf}\\
A_{b\bar\mu} (\omega )& = &\sum _{m,n\geq 0} \mbox{e}^{-\beta E_m^0}
\mid \langle 1,n | b_{\bar\mu}^{\dag} | 0, m\rangle \mid^2
\delta (\omega -(E_n^1 - E_m^0 ))\ ,
\label{lehmannb}
\eqalabel{lehmann}
\end{eqlettarray}
where $E_n^Q$ are the
energy eigenvalues ($E_0^0 \leq E_n^Q$ is the ground state energy) and 
$|Q,n\rangle $ the many--body eigenstates of $H$ in the sector $Q$ of 
Fock space. At zero temperature, $A_f$ reduces to
$A_{f\sigma} (\omega ) = \sum _{n\geq 0} \mid 
\langle 1,n | f_{\sigma}^{\dag} | 0,0\rangle \mid^2
\delta (\omega -( E_n^1 - E_m^0 ))$ and similar for $A_b$. 
It is seen that the $A_{f,b}$ have threshold behavior at 
$\omega = E_0 \equiv E_0^1 - E_0^0$, 
with $A_{f,b}(\omega ) \equiv 0$ for $\omega < E_0$, $T=0$. 
The vanishing imaginary part at frequencies $\omega < 0$ may be
shown to be a general property of all quantities involving slave particle
operators, e.g. also of auxiliary particle self--energies and
vertex functions.

(2) --- As will be seen in section 4.2 (Eq. (\ref{gdNCA})), 
physical expectation values not only involve the particle--like
auxiliary propagators Eq. (\ref{eq:def_gfb1}) but also hole--like
contributions. It is, therefore, useful to define the ``anti--fermion'' 
and ``anti--boson'' propagators (in imaginary time representation)
\begin{eqlettarray}
G^-_{f\sigma}(\tau_1-\tau_2) &=& \Theta (\tau_2-\tau_1)\lim 
_{\lambda\to \infty}
\langle f^{\dag}_{\sigma} (\tau_2) f_{\sigma} (\tau_1) \rangle _G\ \\
G^-_{b\bar\mu}(\tau_1-\tau_2) &=& \Theta (\tau_2-\tau_1)\lim 
_{\lambda\to \infty}
\langle b^{\dag}_{\bar\mu} (\tau_2) b_{\bar\mu} (\tau_1)\rangle _G\ ,
\end{eqlettarray}
whose spectral functions have the Lehmann representations
\begin{eqlettarray}
A^{-}_{f\sigma} (\omega )&=&\sum _{m,n\geq 0} \mbox{e}^{-\beta E_m^1}
\mid \langle 0,n | f_{\sigma} |
1,m\rangle \mid^2\delta (\omega -( E_n^0 - E_m^1)) 
\label{lehmannf-}\\
A^{-}_{b\bar\mu} (\omega )&=&\sum _{m,n \geq 0} \mbox{e}^{-\beta E_m^1}
\mid \langle 0,n | b_{\bar\mu} |
1,m\rangle \mid^2\delta (\omega - ( E_n^0 - E_m^1 ))\ .
\label{lehmannb-}
\eqalabel{lehmann-}
\end{eqlettarray}
$E_0^1$ is the ground state energy in the $Q=1$ sector. The
expressions (\ref{lehmann}) and (\ref{lehmann-}) immediately imply
a relation between $A_{f,b}$ and $A^-_{f,b}$,
\begin{equation}
A^-_{f,b} (\omega ) = \mbox{e}^{-\beta \omega } A_{f,b}(\omega ) \ .
\end{equation}

(3) --- The property of only forward--in--time propagation 
(Eqs. (\ref{eq:def_gfb1}), (\ref{lehmann})) means
that the auxiliary particle propagators $G_{f,b}$
are formally identical to the core propagators of the well--known
X--ray threshold problem \cite{anderson.67}--\cite{mahan.80}. Thus, the 
knowledge of the infrared behavior of the latter may be directly 
applied to the former. In particular, the spectral functions are 
found (see below) to diverge at the threshold $E_0$ in  
a power law fashion (infrared singularity)
\begin{equation}
A_{f,b}(\omega ) \sim | \omega - E_0 |
^{-\alpha_{f,b}}\theta (\omega - E_0)
\label{powerlaw}
\end{equation}
due to a diverging number of particle--hole excitation processes in the
conduction electron sea as $\omega \to E_o$. 

For the single channel case $(M=1)$, i.e. the usual Kondo or mixed
valence problem, the exponents $\alpha_f$ and $\alpha_b$ can be found
analytically from the following chain of arguments:  Anticipating that 
in this case the impurity spin is completely screened by the conduction
electrons at temperature $T = 0$, leaving a pure-potential scattering
center, the ground state $| 1,0 \rangle$ is a slater determinant of 
one--particle scattering states, characterized by scattering phase 
shifts $\eta_\sigma$ in the s--wave channel (assuming for simplicity a
momentum independent hybridization matrix element $V$).
To calculate the fermion spectral function $A_{f\sigma}(\omega)$ at $T=0$ 
from Eq.~(\ref{lehmannf}), one needs to 
evaluate $\langle 1,n\mid f_\sigma^{\dag}\mid 0,0 \rangle$, which
is just the overlap of two slater determinants, an eigenstate of the 
fully interacting Kondo system, $|1,n\rangle$, on the one hand, and the 
ground state of the conduction electron system in the absence of the
impurity combined with the decoupled impurity level occupied by an electron
with spin $\sigma$, $f^{\dag}_{\sigma}|0,0\rangle$, on the other hand.  
As shown by Anderson \cite{anderson.67}, 
the overlap of the two ground state slater determinants,
$\langle 1,0\mid f_\sigma^{\dag}\mid 0,0 \rangle$, tends to zero in
the thermodynamic limit (orthogonality catastrophe). 
Analogous relations hold for the boson spectral function 
$A_{b}(\omega)$. As a result, the long--time relaxation into the 
interacting ground state is inhibited, leading to the infrared power 
law divergence of the spectral functions, Eq. (\ref{powerlaw}).

The X--ray threshold exponents can be expressed in terms of the 
scattering phase shifts at the Fermi level by the exact relation 
\cite{schotte.69}
\begin{eqlettarray}
\alpha_{f,b} &=& 
1 - \sum_{\sigma '}\Big(\frac{\eta_{f,b\;\sigma '}}{\pi}\Big)^2\ .
\end{eqlettarray}
Here the $\eta_{f\sigma '}$ ($\eta_{b\sigma '}$) are the scattering 
phase shifts of the single--particle wave functions in channel $\sigma '$
of the fully interacting ground state $| 1,0\rangle$,
relative to the wave functions of the free state $f_\sigma^{\dag}| 
0,0\rangle$ (\,$b^{\dag}| 0,0\rangle$\,). 
Via the Friedel sum rule, the scattering phase shifts are, in turn, 
related to the change $\Delta n_{c\sigma '}$ of the average number of 
conduction electrons per scattering channel $\sigma '$ due to the 
presence of the impurity: 
$\eta_{f,b\; \sigma '}=  \pi \Delta n_{c\sigma '}$. Obviously, 
$\Delta n_{c\sigma '}$ is equal and opposite in sign to the difference 
of the average impurity occupation numbers of the states  $| 1,0\rangle$ 
and $f_\sigma^{\dag}| 0,0\rangle$ (\,$b^{\dag}| 0,0\rangle$\,). 
Thus, in the pseudofermion propagator $G_{f\sigma }$ we have the 
phase shifts,
\begin{eqlettarray}
\eta_{f\sigma '}= - \pi \Big(\frac{n_d}{N}-\delta _{\sigma\sigma '}\Big) 
\end{eqlettarray}
and in the slave boson propagator $G_b$,
\begin{eqlettarray}
\eta_{b}= - \pi\frac{n_d}{N}\ , 
\end{eqlettarray}
where $n_d$ denotes the total occupation number of the impurity level 
in the interacting ground state. (The term $\delta _{\sigma\sigma '}$ in
$\eta_{f\; \sigma '}$ appears because  $f_\sigma^{\dag}| 0,0\rangle$ 
has impurity occupation number 1.)
For example, in the Kondo limit $n_d \rightarrow 1$ and for a 
spin $1/2$ impurity 
($N=2$) this leads to resonance scattering, $\eta_{f,b\;\sigma '} = 
\pi/2$. As a result, one finds \cite{mengemuha.88} for the threshold 
exponents
\begin{eqlettarray}
\alpha_f &=& \frac{2n_d - n_d^2}{N}\\
\alpha_b &=& 1-\frac{n_d^2}{N}
\end{eqlettarray}
These results have been found independently from Wilson's numerical
renormalization group approach \cite{costi.94,costi.94b} and using the Bethe 
ansatz solution and boundary conformal field theory \cite{fujimoto.96}.  
It is interesting to note that (i) the exponents depend on the 
level occupancy $n_d$ 
(in the Kondo limit $n_d \rightarrow 1$, $\alpha_f = 1/N$ and
$\alpha_b = 1 - 1/N$, whereas in the opposite, empty orbital, limit
$n_d \rightarrow 0$, $\alpha_f \rightarrow 0$ and $\alpha_b
\rightarrow 1$) (ii) the sum of the exponents $\alpha_f + \alpha_b = 1
+ 2 \frac{n_d(1-n_d)}{N} \geq  1$.

We stress that the above derivation of the infrared exponents
$\alpha_{f,b}$ holds true only if the impurity complex acts as a pure
potential scattering center at $T = 0$.  This is equivalent to the
statement that the conduction electrons behave locally, i.e. at the
impurity site, like a Fermi liquid. Conversely, in the multi--channel 
(non--FL) case, $N\geq 2$, $M\geq N$, the exponents have been found 
from a conformal field theory solution \cite{affleck.91} of the problem
in the Kondo limit to be $\alpha _f = M/(M+N)$, $\alpha _b = N/(M+N)$,
which differ from the FL values. Thus, one may infer from
the values of $\alpha_{f,b}$ as a function of $n_d$, whether or not
the system is in a local Fermi liquid state.

\section{Mean field and non-crossing approximations}
For physical situations of interest, the $s-d$ hybridization of the
Anderson model (\ref{Ahamilton}) is much smaller than the conduction 
band width, ${\cal N}(0) V \ll 1$, where ${\cal N}(0) =1/D$ 
is the local conduction 
electron density of states at the Fermi level.  This suggests a 
perturbation expansion in ${\cal N}(0)V$. A straightforward expansion in terms 
of bare Green's functions is not adequate, as it would not allow to 
capture the physics of the Kondo screened state, or else the infrared
divergencies of the auxiliary particle spectral functions discussed in
the last section.  In the framework of the slave boson representation,
two types of nonperturbative approaches have been developed.  The
first one is mean field theory for both the slave boson amplitude 
$\langle b\rangle$ and the constraint 
($\langle Q\rangle = 1$ rather than $Q = 1$).  The second one
is resummation of the perturbation theory to infinite order.

\subsection{Slave boson mean field theory} 
Slave boson mean field theory is based on the assumption that the slave
bosons condense at low temperatures such that $\langle b_{\bar\mu}\rangle\neq 0$.
Replacing the operator $b_{\bar\mu}$ in $H + \lambda Q$  
by $\langle b_{\bar\mu} \rangle$ (see Ref.~\cite{readnewns.88}), 
where $\lambda$ is a Lagrange multiplier to be adjusted such that
$\langle Q \rangle = 1$, one arrives at a resonance level model for the
pseudofermions.  The position of the resonance, $E_d + \lambda$, is
found to be given by the Kondo temperature $T_K$, and is thus close to
the Fermi energy. The resonance generates the low energy scale $T_K$,
and leads to local Fermi liquid behavior.  While this is qualitatively
correct in the single--channel case, it is in blatant disagreement with
the exactly known behavior in the multi--channel case.  The mean field
theory can be shown to be exact for $M=1$ in the limit $N \rightarrow
\infty$ for a model in which the constraint is softened to be $Q=N/2$. 
However, for finite $N$ the breaking of the local gauge symmetry,
which would be implied by the condensation of the slave boson field,
is forbidden by Elitzur's theorem \cite{elitzur.75}.
It is known that for finite $N$ the fluctuations 
in the phase of the complex expectation value $\langle b_{\bar\mu}\rangle$ 
are divergent and lead to the suppression of $\langle b_{\bar\mu} \rangle$ to 
zero (see also \cite{jevicki.77}--\cite{lawrie.85}). 
This is true in the cartesian
gauge, whereas in the radial gauge the phase fluctuations may be shown
to cancel at least in lowest order. It has not been possible
to connect the mean field solution, an apparently reasonable
description at low temperatures and for $M = 1$, to the high 
temperature behavior $(T \gg T_K)$, dominated by logarithmic temperature 
dependence, in a continuous way \cite{readnewns.88}.  
Therefore, it seems that the slave boson mean field
solution does not offer a good starting point even for only a
qualitatively correct description of quantum impurity models.

\subsection{1/N expansion vs. self--consistent formulation}
The critical judgement of mean field theory is corroborated by the
results of a straightforward $1/N$-expansion in the single channel
case, keeping the exact constraint, and not allowing for a finite
bose field expectation value \cite{kuroda.88}.  
Within this scheme the exact behavior of the thermodynamic quantities 
(known from the Bethe ansatz solution) at low temperatures as well as 
high temperatures is recovered to the considered order in $1/N$.  
Also, the exact auxiliary particle exponents $\alpha _{f,b}$ are 
reproduced in order $1/N$, using a plausible exponentiation 
scheme \cite{kuroda.97}.

In addition, dynamical quantities like the d-electron spectral
function and transport coefficients can be calculated exactly to a
desired order in $1/N$, within this approach.  However, as clear-cut
and economical this method may be, it does have serious limitations.
For once, the experimentally most relevant case of $N=2$ or somewhat
larger is not accessible in $1/N$ expansion. Secondly, non-Fermi
liquid behavior, being necessarily non-perturbative in $1/N$, cannot
be dealt with in a controlled way on the basis of a $1/N$-expansion.
To access these latter two regimes, a new approach non-perturbative 
in $1/N$ is necessary.
\begin{figure}
\vspace*{-0cm}
\centerline{\psfig{figure=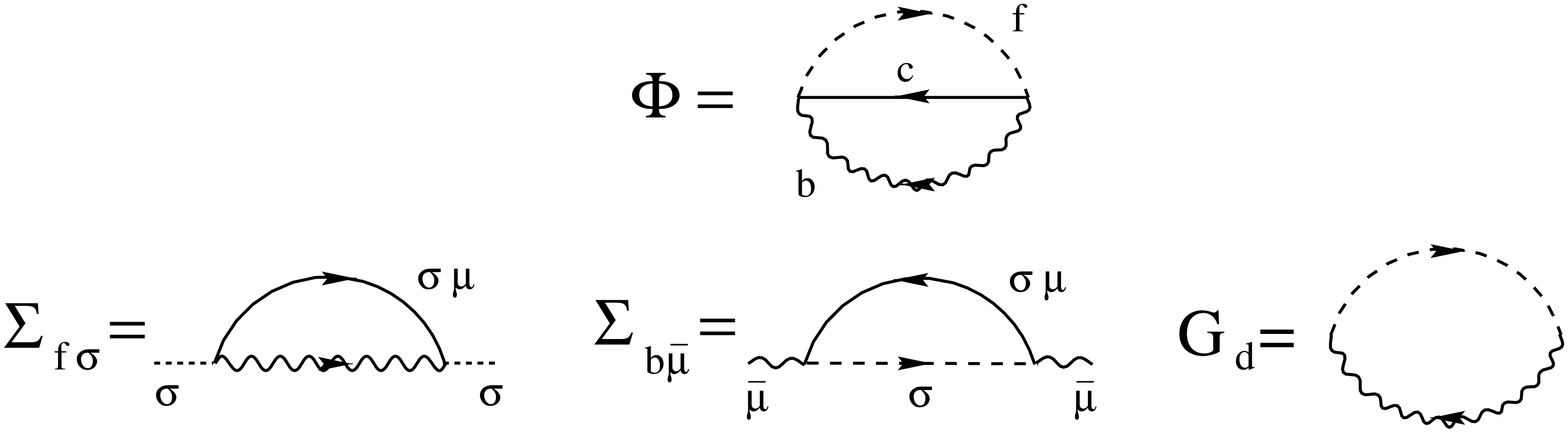,width=\textwidth}}
\vspace*{0.6cm}
\caption{
Diagrammatic representation of the generating functional $\Phi $ 
of the NCA. Also shown are the pseudoparticle self--energies and the
local electron Green's function derived from $\Phi$, Eqs.~(19)--(21).
Throughout this article, dashed, wavy and solid lines represent fermion, 
boson, and conduction electron lines, respectively. In the diagram for 
$\Sigma _{f\sigma}$ the spin labels are shown explicitly to 
demonstrate that there are no coherent spin fluctuations taken into account.
\label{NCA}}
\end{figure}

We conjecture that this new approach is gauge invariant many-body
theory of pseudofermions and slave bosons. As long as gauge symmetry
violating objects such as Bose field expectation values or fermion
pair correlation functions do not appear in the theory, gauge
invariance of physical quantities can be guaranteed in suitably chosen
approximations by the proper match of pseudofermion and slave boson
properties, without introducing an additional gauge field.  This
requires the use of conserving approximations \cite{kadanoff.61}, 
derived from a Luttinger-Ward functional $\Phi$.  
$\Phi$ consists of all vacuum
skeleton diagrams built out of fully renormalized Green's functions
$G_{b,f,c}$ and the bare vertex $V$.  The self--energies
$\Sigma_{b,f,c}$ are obtained by taking the functional derivative of
$\Phi$ with respect to the corresponding Green's function (cutting the
Green's function line in each diagram in all possible ways),
\begin{equation}
\Sigma_{b,f,c} = \delta \Phi/\delta G_{b,f,c}.
\label{fderiv}
\end{equation}
Irreducible vertex functions, figuring as integral kernels in
two-particle Bethe-Salpeter equations, are generated by second order
derivatives of $\Phi$.

The choice of diagrams for $\Phi$ defines a given approximation. It
should be dictated by the dominant physical processes and by
expansion in a small parameter, if available.  As noted before, in the
present context, we may take the hybridization $V$ to be a small
quantity (dimensionless parameter $N_oV$).  This suggests to start
with the lowest order (in $V$) diagram of $\Phi$, which is second
order (see Fig.~\ref{NCA}).  The self--energies generated from this 
obey after analytic continuation to real frequencies 
($i\omega \rightarrow \omega -i0$) and projection the following 
equations of self--consistent second order perturbation theory 
\begin{eqlettarray}
\Sigma_{f\sigma}^{(NCA)}(\omega-i0)&=&\Gamma\sum _{\mu}\int 
              \frac{{d}\varepsilon}{\pi}\,
               [1-f(\varepsilon )]
              A_{c\mu\sigma}^0(\varepsilon)G_{b\bar\mu}(\omega -\varepsilon -i0)
              \label{sigfNCA}\\
\Sigma_{b\bar\mu}^{(NCA)}(\omega -i0)&=&\Gamma\sum _{\sigma}\int 
              \frac{{d}\varepsilon}{\pi}\,
              f(\varepsilon )A_{c\mu\sigma}^0(\varepsilon)
              G_{f\sigma}(\omega +\varepsilon -i0)
              \label{sigbNCA}\\
G_{d\mu\sigma}^{(NCA)}(\omega -i0)
         &=& \int  {d}\varepsilon\,  {\rm e}^{-\beta\varepsilon}
         [ G_{f\sigma}(\omega +\varepsilon -i0)A_{b\bar\mu}(\varepsilon )
          \nonumber\\
         &\ &\hspace*{1.8cm}-A_{f\sigma}(\varepsilon )
                   G_{b\bar\mu}(\varepsilon -\omega +i0) ]\\
         &=& \int  {d}\varepsilon\,  
         [ G_{f\sigma}(\omega +\varepsilon -i0)A^-_{b\bar\mu}(\varepsilon )
          -A^-_{f\sigma}(\varepsilon )
                   G_{b\bar\mu}(\varepsilon -\omega +i0) ]\; , \nonumber
              \label{gdNCA}
\eqalabel{eq:NCA}
\end{eqlettarray}
where $A_{c\mu\sigma}^0=\frac{1}{\pi}\, 
{\rm Im}G_{c\mu\sigma}^0/{\cal N}(0)$ is the
(free) conduction electron density of states per spin and channel,
normalized to the density of states at the Fermi level ${\cal N}(0)$, and 
$f(\varepsilon )=1/({\rm exp}(\beta\varepsilon )+1)$ denotes the 
Fermi distribution function. Together with the expressions 
(\ref{green}), (\ref{green0}) for the Green's functions,
Eqs. (\ref{sigfNCA})--(\ref{gdNCA}) form a set of self--consistent 
equations for $\Sigma_{b,f,c}$, comprised of all diagrams without 
any crossing propagator lines and are, thus, known as the
``non--crossing approximation'', in short NCA \cite{keiter.71,kuramoto.83}.

At zero temperature and for low frequencies Eqs. (\ref{sigfNCA}) 
and (\ref{sigbNCA}) may be converted into a set of linear differential 
equations for $G_f$ and $G_b$ \cite{muha.84}, 
which allow to find the infrared exponents as $\alpha_f =
\frac{M}{M+N}$; $\alpha_b = \frac{N}{M+N}$, independent of $n_d$.  For
the single channel case these exponents do not agree with the exact
exponents derived in section 3. This indicates that the NCA is
not capable of recovering the local Fermi liquid behavior for $M=1$.
A numerical evaluation of the $d$-electron Green's function, which 
is given by the local self--energy $\Sigma_c$ divided by $V^2$ and 
hence is given by the boson--fermion bubble within NCA (Fig.~\ref{NCA}), 
shows indeed a spurious singularity at the Fermi energy \cite{costi.96}.  
The NCA performs somewhat better in the multi--channel case, 
where the exponents $\alpha_f$ and $\alpha_b$ yield the correct 
non--Fermi liquid exponents
of physical quantities as known from the Bethe ansatz 
solution \cite{wiegmann.83} and conformal field 
theory \cite{affleck.91}. However, the specific heat and
the residual entropy are not given correctly in NCA.  Also, the
limiting low temperature scaling laws for the thermodynamic
quantities are attained only at temperatures substantially below
$T_K$, in disagreement with the exact Bethe ansatz solution.

\subsection{Low--temperature evaluation of the self--consistency 
equations}
In order to enter the asymptotic power law regime of the auxiliary 
spectral functions the self--consistent scheme, in particular the NCA, 
must be evaluated for temperatures several orders of magnitude below $T_K$, 
the low temperature scale of the model. The equations are solved 
numerically by iteration. In the following we describe the two main 
procedures to make the diagrammatic auxiliary particle technique 
suitable for the lowest temperatures.

The grand--canonical expectation value of the auxiliary particle 
number appearing in Eq.~(\ref{eq:GdNCA}) is given in terms of the 
grand--canonical (unprojected) auxiliary particle spectral functions 
${\cal A}_{f,b}(\omega,\lambda)$ by,
\begin{eqnarray}
\langle Q \rangle _{G} (\lambda ) = 
\int  {d}\omega
\Big[ f(\omega) \sum _{\sigma } {\cal A}_{f\sigma } (\omega,\lambda)
+ b(\omega) \sum _{\mu } {\cal A}_{b\bar\mu}(\omega,\lambda)
\Big]
\end{eqnarray}
where $f(\omega)$, $b(\omega)$ are the Fermi and Bose distribution 
functions, respectively. 
Substituting this into the expression (\ref{eq:zcan}) for the 
canonical partition function we obtain after carrying out
the transformation $\omega \rightarrow \omega +\lambda $, and taking
the limit $\lambda \rightarrow \infty $
\begin{eqnarray}
e^{-\beta F_{imp}(T)} & \equiv & 
\frac{Z_{C}}{Z_{Q=0}}  = 
\lim_{\lambda\rightarrow\infty}e^{\beta\lambda}
\langle Q \rangle_{G}(\lambda) \nonumber\\
& = & \int {d}\omega e^{-\beta\omega} \Big[  \sum _{\sigma } 
A_{f\sigma }(\omega) +\sum _{\mu} A_{b\bar\mu}(\omega)\Big]\ .
\label{eq:Qproj}
\end{eqnarray}
By definition 
$F_{imp}=-\frac{1}{\beta}\ln(Z_{C}/Z_{Q=0})$ is the impurity 
contribution to the Free energy.

The numerical evaluation of expectation values like 
$\langle Q\rangle _{G}(\lambda\rightarrow\infty)$ (Eq. (\ref{eq:Qproj}))
or $G_{d\mu\sigma}(\omega,\lambda\rightarrow\infty)$ (Eq. (\ref{gdNCA})) 
is non--trivial, (1) because at $T=0$ the auxiliary spectral functions 
$A_{f,b}(\omega,T)$
are divergent at the threshold frequency $E_{0}$, where the exact
position of $E_{0}$ is a priori not known, and (2) because 
the Boltzmann factors $e^{-\beta\omega}$ diverge strongly for 
$\omega < 0$. Therefore, we apply the following transformations:

(1) Before performing the projection $\omega \rightarrow \omega +\lambda$,
$\lambda\rightarrow\infty$ we re--define the frequency scale of all
auxiliary particle functions $A_{f,b}$ according to  
$\omega \rightarrow \omega +\lambda_0$, where $\lambda_0$ is a finite
parameter. In each iteration $\lambda _0$  is then determined  such that
\begin{equation}
\int {d}\omega
e^{-\beta\omega} \Big[ \sum _{\sigma } A_{f\sigma }(\omega) +
\sum _{\mu} A_{b\bar\mu}(\omega)\Big]=1
\label{eq:fixl}
\end{equation}
where $A_{f,b}(\omega)=\lim_{\lambda\rightarrow\infty}
A_{f,b}(\omega+\lambda_0,\lambda)$ 
is now an auxiliary spectral function with the new reference energy.
It is seen by comparison with Eq.~(\ref{eq:Qproj}) that 
$\lambda_0(T)=F_{imp}(T) = F_{Q=1}(T)-F_{Q=0}(T)$, i.e. $\lambda_0$ 
is the chemical potential for the auxiliary particle number $Q$, or 
equivalently the impurity contribution to the Free energy. 
The difference of the Free energies becomes equal to the threshold energy 
$E_{0}=E_{Q=1}^{GS}-E_{Q=0}^{GS}$ at $T=0$. More importantly, however, 
the above way of determining a ``threshold'' is less {\em ad hoc}\/ than,
for example, defining it by a maximum in some function appearing in the 
NCA equations. It is also seen from Eq.~(\ref{eq:fixl}) that this procedure 
defines the frequency scale of the auxiliary particles such that 
the $T=0$ threshold divergence of the
spectral functions is at the {\it fixed} frequency $\omega = 0$.
This substantially increases the precision as well as the speed of 
numerical evaluations. Eq. (\ref{eq:GdNCA}) for the projected 
$d$--electron Green's function becomes
\begin{equation}
G_d(\omega ) = \lim _{\lambda \rightarrow \infty} e^{\beta\lambda}
G_{d}(\omega,T,\lambda ).
\end{equation}
  
(2) The divergence of the Boltzmann factors implies that the 
self--con\-sis\-tent solutions for $A_{f,b} (\omega)$
vanish exponentially $\sim e^{\beta\omega}$ for negative frequencies,
confirming their threshold behavior.
It is convenient, not to formulate the self--consistent equations in 
terms of $A_{f,b}$ like in earlier evaluations \cite{bickers.87}, 
but to define new functions $\tilde A_{f,b}(\omega)$ and 
$\mbox{Im}\tilde\Sigma_{f,b}(\omega)$ such that
\begin{eqnarray}
A_{f,b} (\omega) & = & f(-\omega )~ \tilde A_{f,b}(\omega)\\
\mbox{Im}\Sigma_{f,b}(\omega) & = & f(-\omega ) ~ \mbox{Im} 
\tilde\Sigma_{f,b}(\omega).
\end{eqnarray}
After fixing the chemical potential $\lambda _0$ and performing the
projection onto the physical subspace, the canonical partition function
(Eq.~(\ref{eq:zcan})) behaves as 
$\lim _{\lambda\rightarrow\infty}e^{\beta(\lambda-
\lambda_0)}\;Z_C(T) =1$. 
In this way all exponential divergencies are absorbed by one single 
function for each particle species.
The NCA equations in terms of these functions are free of divergencies 
of the statistical factors and read
\begin{eqnarray}
\mbox{Im}\tilde\Sigma_{f\sigma}(\omega -i0) 
& = & \Gamma\sum_{\mu} \int {d}\varepsilon\,
\frac{f(-\varepsilon)(1-f(\omega -\varepsilon))}{1-f(\omega )} 
A_{c\mu\sigma}^0(\varepsilon)
\tilde A_{b\bar\mu}(\omega-\varepsilon)\\
\mbox{Im}\tilde\Sigma_{b\bar\mu}(\omega -i0) 
& = & \Gamma\sum_{\sigma} \int {d}\varepsilon\,
\frac{f(\varepsilon)(1-f(\omega +\varepsilon))}{1-f(\omega )} 
A_{c\mu\sigma}^0(\varepsilon)
\tilde A_{f\sigma}(\omega+\varepsilon)\\
\langle Q \rangle  (\lambda _0,\lambda\rightarrow\infty) & = &
\int {d}\omega
f(\omega)\Big[ \sum _{\sigma } \tilde A_{f\sigma }(\omega) +
\sum _{\mu}\tilde A_{b\bar\mu} (\omega)\Big] = 1\\
\mbox{Im}G_{d\sigma}(\omega -i0) & = & 
\int {d}\varepsilon [f(\varepsilon +\omega)f(-\varepsilon )+
f(-\varepsilon -\omega)f(\varepsilon)]
\tilde A_{f\sigma}(\varepsilon +\omega)\tilde 
A_b(\varepsilon)\nonumber.\\
\end{eqnarray}
The real parts of the self--energies $\Sigma _f$, $\Sigma _b$ 
are determined from $\mbox{Im}\Sigma _f$, $\mbox{Im}\Sigma _b$ 
through a Kramers--Kroenig relation, and the auxiliary functions
$\tilde A_{f\sigma}(\omega)=\frac{1}{\pi}
\mbox{Im}\tilde\Sigma _{f\sigma}(\omega -i0)/
\bigl[\bigl(\omega+\lambda_0-i0-E_d-\mbox{Re}\Sigma_{f\sigma}
(\omega-i0)\bigr)^2+\mbox{Im}\Sigma_{f\sigma}(\omega-i0)^2\bigr]$, 
$\tilde A_{b\bar\mu}(\omega)=\frac{1}{\pi}
\mbox{Im}\tilde\Sigma _{b\bar\mu}(\omega -i0)/
\bigl[\bigl(\omega+\lambda_0-i0-\mbox{Re}\Sigma_{b\bar\mu}
(\omega-i0)\bigr)^2+\mbox{Im}\Sigma_{b\bar\mu}(\omega-i0)^2\bigr]$,
thus closing the above set of equations. 

The method described above allows to solve the NCA equations 
effectively for temperatures down to typically $T=10^{-4}T_K$. 
It may be shown that the same procedure can also be applied to 
self--consistently compute vertex corrections beyond the NCA 
(see section 5).

\section{Conserving T-matrix approximation}
\subsection{Dominant contributions at low energy}
\begin{figure}
\vspace*{-0cm}
\centerline{\psfig{figure=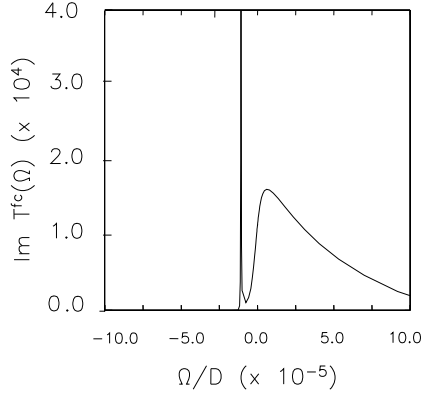,width=5.5cm}}
\vspace*{0.6cm}
\caption{ 
Imaginary part of the conduction electron--pseudofermion 
T--matrix $T^{(cf)}$ as a function of the
COM frequency $\Omega$ for the single--channel case $M=1$, $N=2$, 
evaluated by inserting NCA solutions for the
intermediate state propagators ($E_d=-0.67 D$, $\Gamma = 0.15 D$,
$T=4\cdot 10^{-3}T_K$).
The contribution from the pole positioned at a negative frequency
$\Omega = \Omega _{cf} \simeq -T_K$ (compare text) is clearly seen.
\label{tmatpole0}}
\end{figure}

In order to eliminate the shortcomings of the NCA mentioned above, 
the guiding principle should be to find 
contributions to the vertex functions which renormalize the
auxiliary particle threshold exponents to their correct values,
since this is a necessary condition for the description of 
FL and non--FL behavior, as discussed in section 3. Furthermore,
it is instructive to realize that in NCA any coherent spin flip
and charge transfer processes are neglected, as can be seen 
explicitly from Eqs.~(\ref{sigfNCA}), (\ref{sigbNCA}) or from 
Fig.~\ref{NCA}. These processes are known to be responsible for the
quantum coherent collective behavior of the Anderson impurity
complex below $T_K$. The existence of collective excitations
in general is reflected in a singular behavior of the 
corresponding two--particle vertex functions. In view of the
tendency of Kondo systems to form a collective spin singlet
state, we are here interested in the spin singlet channel
of the pseudofermion--conduction electron vertex function and
in the slave boson--conduction electron vertex function.
It may be shown by power counting arguments (compare appendix A)
that there are no corrections to the NCA exponents in any finite 
order of perturbation theory \cite{coxruck.93}. Thus, we 
are led to search for singularities in the aforementioned vertex 
functions arising from an infinite resummation of terms.   
\begin{figure}
\vspace*{-0cm}
\centerline{\psfig{figure=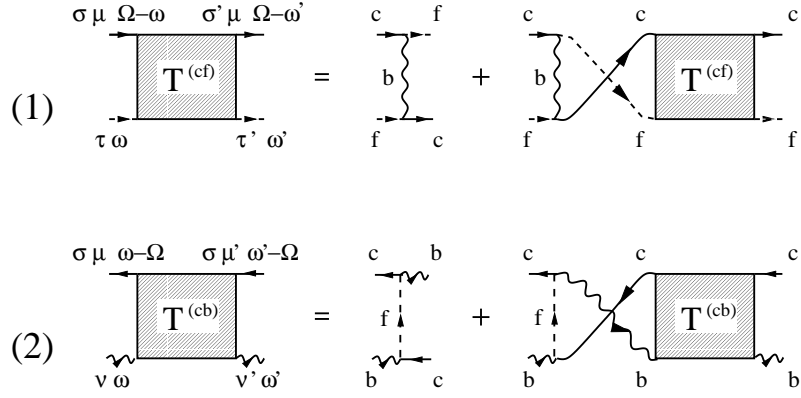,width=10.5cm}}
\vspace*{0.6cm}
\caption{
Diagrammatic representation of the Bethe--Salpeter equation for (1) 
the conduction electron--pseudofermion
T-matrix $T^{(cf)}$, Eq.~(\ref{cftmateq}), and (2) the conduction 
electron--slave boson T-matrix $T^{(cb)}$, Eq.~(\ref{cbtmateq}). 
$T^{(cb)}$ is obtained from $T^{(cf)}$ by 
interchanging $f \leftrightarrow b$ and $c\leftrightarrow c^{\dag}$.
\label{tmat}}
\end{figure}
From the preceding discussion it is natural to perform a partial 
resummation of those terms which, at each order in the 
hybridization $V$, contain the maximum number of spin flip or charge 
fluctuation processes, respectively. This amounts to calculating the 
conduction electron--pseudofermion vertex function in
the ``ladder'' approximation definied in Fig.~\ref{tmat}, where the 
irreducible vertex is given by $V^2G_b$. In analogy to similar 
resummations for an interacting one--component Fermi system, we 
call the total c--f vertex function T--matrix $T^{(cf)}$. 
The Bethe--Salpeter equation for $T^{(cf)}$ reads (Fig.~\ref{tmat} (1)),
\begin{eqnarray}
T^{(cf)\ \mu}_{\sigma\tau,\sigma '\tau '}
(i\omega _n, i\omega _n ', i\Omega _n ) =
&+&V^2G_{b\bar\mu}(i\omega _n + i\omega _n ' - i\Omega _n ) 
\delta _{\sigma\tau '}\delta _{\tau \sigma '}
\nonumber\\
&-&V^2T\sum _{\omega _n''}G_{b\bar\mu}(i\omega _n + i\omega _n '' - 
i\Omega _n ) \times 
\label{cftmateq}\\
&&\hspace*{-0.6cm}
G_{f\sigma}(i\omega _n'') \ G^0_{c\mu\tau}(i\Omega _n -i\omega _n '')\ 
T^{(cf)\ \mu}_{\tau \sigma,\sigma '\tau '}(i\omega _n '', i\omega _n ', 
i\Omega _n ). \nonumber
\end{eqnarray}\noindent
A similar  integral equation holds for the charge fluctuation T--matrix
$T^{(cb)}$ (Fig.~\ref{tmat} (2)). 
\begin{eqnarray}
T^{(cb)\ \sigma}_{\mu\nu,\mu '\nu '}(i\omega _n,i\omega _n ',i\Omega _n) =
&+&V^2G_{f\sigma}(+i\omega _n + i\omega _n ' - i\Omega _n ) 
\delta _{\mu\nu '}\delta _{\nu \mu ' }
\nonumber\\
&-&V^2T\sum _{\omega _n''}G_{f\sigma}(i\omega _n + i\omega _n '' - 
i\Omega _n ) \times 
\label{cbtmateq}\\
&&\hspace*{-0.6cm}
G_{b\bar\mu}(i\omega _n'') \ G^0_{c\nu\sigma}(-i\omega _n ''-i\Omega _n )\ 
T^{(cb)\ \sigma}_{\nu\mu,\mu '\nu '}(i\omega _n '', i\omega _n ', 
i\Omega _n ). \nonumber
\end{eqnarray}\noindent
In the above Bethe--Salpeter equations $\sigma$, $\tau$, $\sigma '$, 
$\tau '$ represent spin and $\mu$, $\nu$, $\mu '$, $\nu '$ channel indices.
Inserting NCA Green's functions for the intermediate state
propagators of Eq. (\ref{cftmateq}) and solving it numerically, 
we find at low temperatures and in the Kondo regime $(n_d \gsim  0.7)$ 
a pole of $T^{(cf)}$ in the singlet channel (see appendix A) 
as a function of the center--of--mass (COM) frequency $\Omega$, at a 
frequency which scales with the Kondo temperature, 
$\Omega = \Omega_{cf} \simeq - T_K$. 
This is shown in Fig.~\ref{tmatpole0}. The threshold behavior of
the imaginary part of $T^{(cf)}$ as a function of $\Omega$ with
vanishing spectral weight at negative frequencies and temperature 
$T=0$ is clearly seen. In addition, a very sharp structure appears,
whose broadening is found to vanish as the temperature tends to zero, 
indicative of a pole in $T^{(cf)}$ at the {\it real} frequency
$\Omega _{cf}$, i.e.~the tendency to form a collective singlet 
state between the conduction electrons and the localized spin.   
Similarly, the corresponding $T$-matrix $T^{(cb)}$ in the conduction
electron--slave boson channel, evaluated within the analogous 
approximation, develops a pole at negative values of $\Omega$ in the 
empty orbital regime $(n_d \lsim 0.3)$.  In the mixed valence 
regime ($n_d \simeq 0.5)$ the poles in both $T^{(cf)}$ 
and $T^{(cb)}$ coexist. The appearance of poles in the two--particle 
vertex functions $T^{(cf)}$ and $T^{(cb)}$, which signals the formation 
of collective states, may be expected to influence the behavior of 
the system in a major way.

\subsection{Self--consistent formulation: CTMA}
\begin{figure}
\vspace*{-0cm}
\centerline{\psfig{figure=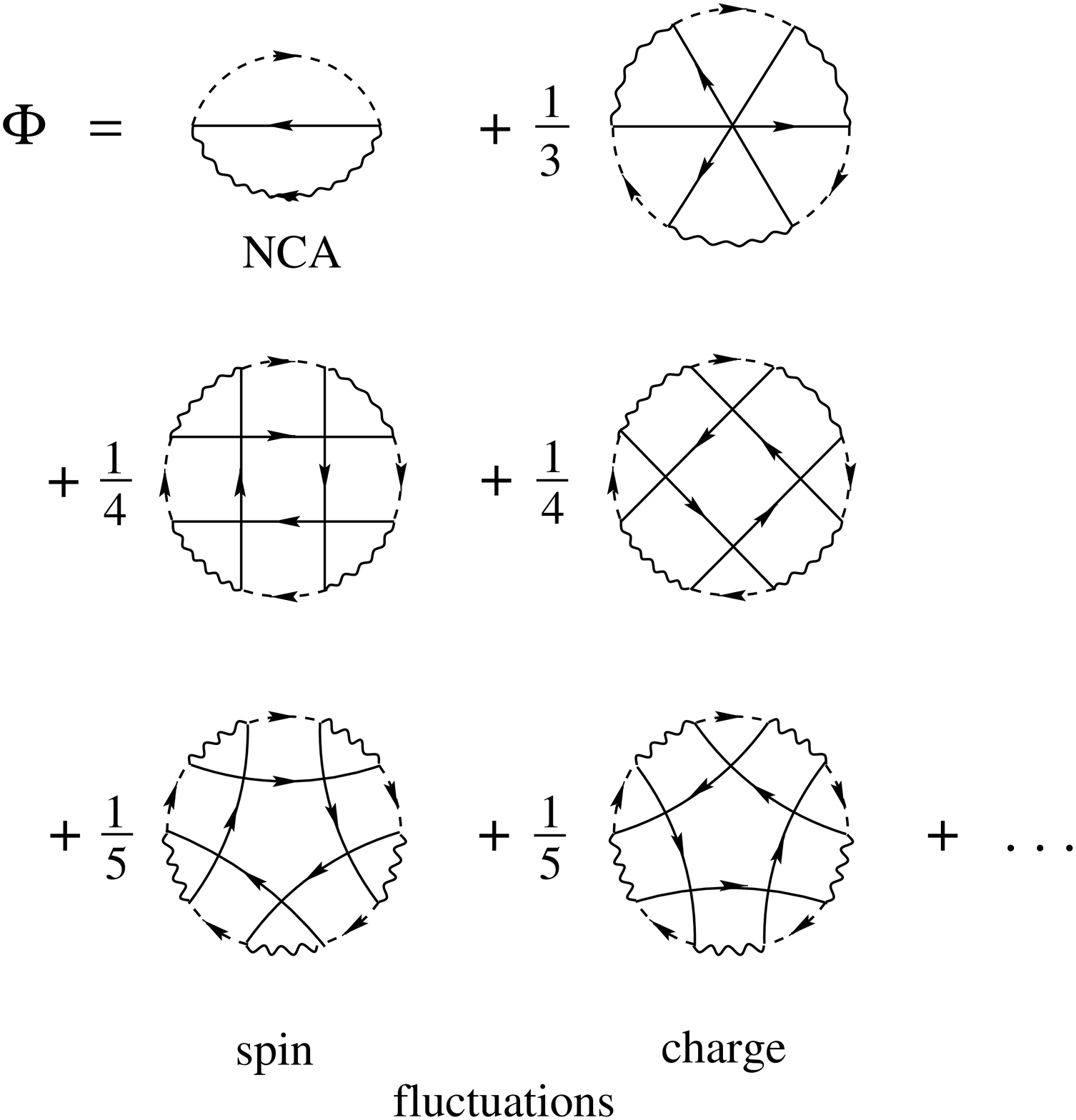,width=9.4cm}}
\vspace*{0.6cm}
\caption{ 
Diagrammatic representation of the
Luttinger--Ward functional generating the conserving
T--matrix approximation (CTMA). The terms with the conduction 
electron lines running clockwise (labelled ``spin fluctuations'') generate 
the T--matrix $T^{(cf)}$, while the terms with the conduction electron 
lines running counter--clockwise (labelled ``charge fluctuations'')
generate the T--matrix $T^{(cb)}$.
\label{CTMA}}
\end{figure}

On the level of approximation considered so far, the description is
not yet consistent: In the limit of zero temperature the spectral
weight of $T^{(cf)}$ and $T^{(cb)}$ at negative frequencies $\Omega$
should be strictly zero (threshold property). Nonvanishing 
spectral weight at $\Omega < 0$ like   
a pole contribution for negative $\Omega$ in
$T^{(cf)}$ or $T^{(cb)}$ would lead to a diverging contribution to the
self--energy, which is unphysical. However, 
recall that a minimum requirement on the approximation used is the
conservation of gauge symmetry. This requirement is not met when the
integral kernel of the $T$-matrix equation is approximated by the NCA
result. Rather, the approximation should be generated
from a Luttinger--Ward functional. The corresponding generating 
functional is shown in Fig.~\ref{CTMA}. It is defined as the 
infinite series of all vacuum skeleton diagrams which consist
of a single ring of auxiliary particle propagators, where each
conduction electron line spans at most two hybridization vertices
(Fig.~\ref{CTMA}).
As shown in appendix A by means of a cancellation theorem, the
CTMA includes, at any given loop order, all infrared singular
contributions to leading and subleading order in the frequency
$\omega$. 
The first diagram of the infinite series of CTMA terms
corresponds to NCA (Fig.~\ref{NCA}). The diagram containing two 
boson lines is excluded since it is not a skeleton. Although the 
spirit of the present theory is different from a large $N$ 
expansion, it should be noted that the sum of the $\Phi$ diagrams 
containing up to four boson lines includes all terms of a $1/N$ 
expansion up to $O(1/N^2)$ \cite{anders.94}. By functional 
differentiation with respect to the conduction electron Green's 
function and the pseudofermion or the slave boson propagator, 
respectively, the shown $\Phi$ functional generates the ladder 
approximations $T^{(cf)}$, $T^{(cb)}$ 
for the total conduction electron--pseudofermion vertex
function (Fig.~\ref{tmat}) and for the total conduction 
electron--slave boson vertex function. The auxiliary particle
self--energies are obtained in the conserving scheme as the 
functional derivatives of $\Phi$ with respect to $G_f$ or
$G_b$, respectively (Eq.~(\ref{fderiv})). This defines a set 
of self--consistency equations, which we term 
conserving T--matrix approximation (CTMA), 
where the self--energies are given as  
nonlinear and nonlocal (in time) functionals of the Green's functions,
while the Green's functions are in turn expressed in terms of the
self--energies. The CTMA equations are derived explicitly in 
appendix B. The solution of these equations 
requires that the T--matrices  have vanishing 
spectral weight at negative COM frequencies $\Omega$. Indeed, the
numerical evaluation shows that the poles of $T^{(cf)}$ and $T^{(cb)}$ 
are shifted to $\Omega = 0$ by self--consistency, where they merge 
with the continuous spectral weight present for $\Omega >0$, thus 
renormalizing the threshold exponents of the auxiliary spectral 
functions.

The self--consistent solutions are obtained by first solving the
linear Bethe--Salpeter equations for the T--matrices
by matrix inversion, computing the auxiliary particle 
self--energies from $T^{(cf)}$ and $T^{(cb)}$, and then constructing
the fermion and boson Green's functions from the respective 
self--energies. This process is iterated until convergence is reached.  
We have obtained reliable solutions  
down to temperatures of the order of at least $10^{-2} T_K$
both for the single-channel and for the two-channel Anderson
model. Note that $T_K\rightarrow 0$ in the Kondo limit; 
in the mixed valence and empty impurity regimes, significantly lower
temperatures may be reached, compared to the low temperature scale 
of the model. 
\begin{figure}
\hspace*{0cm}{\psfig{figure=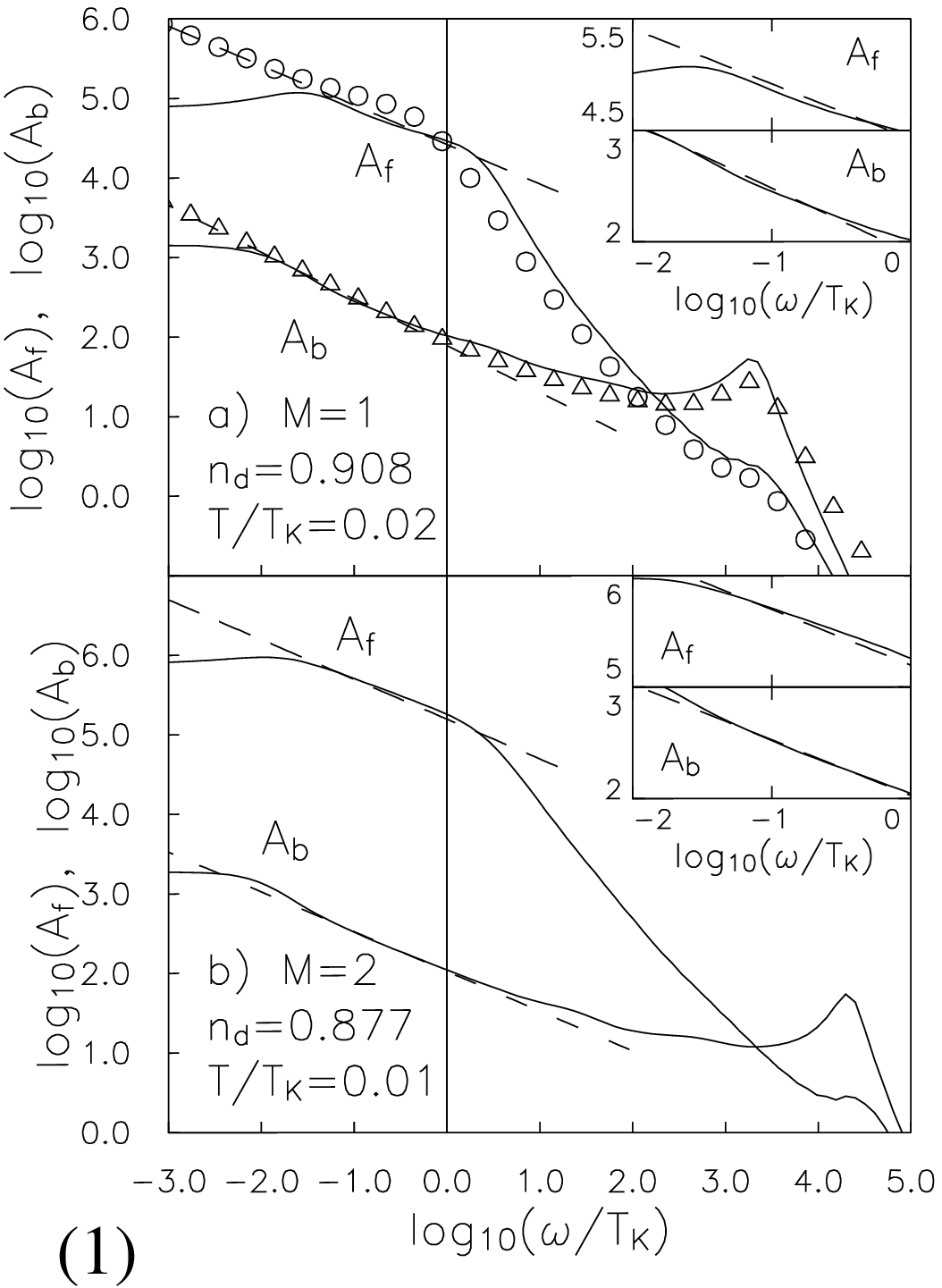,width=6cm}}
\hfill{\psfig{figure=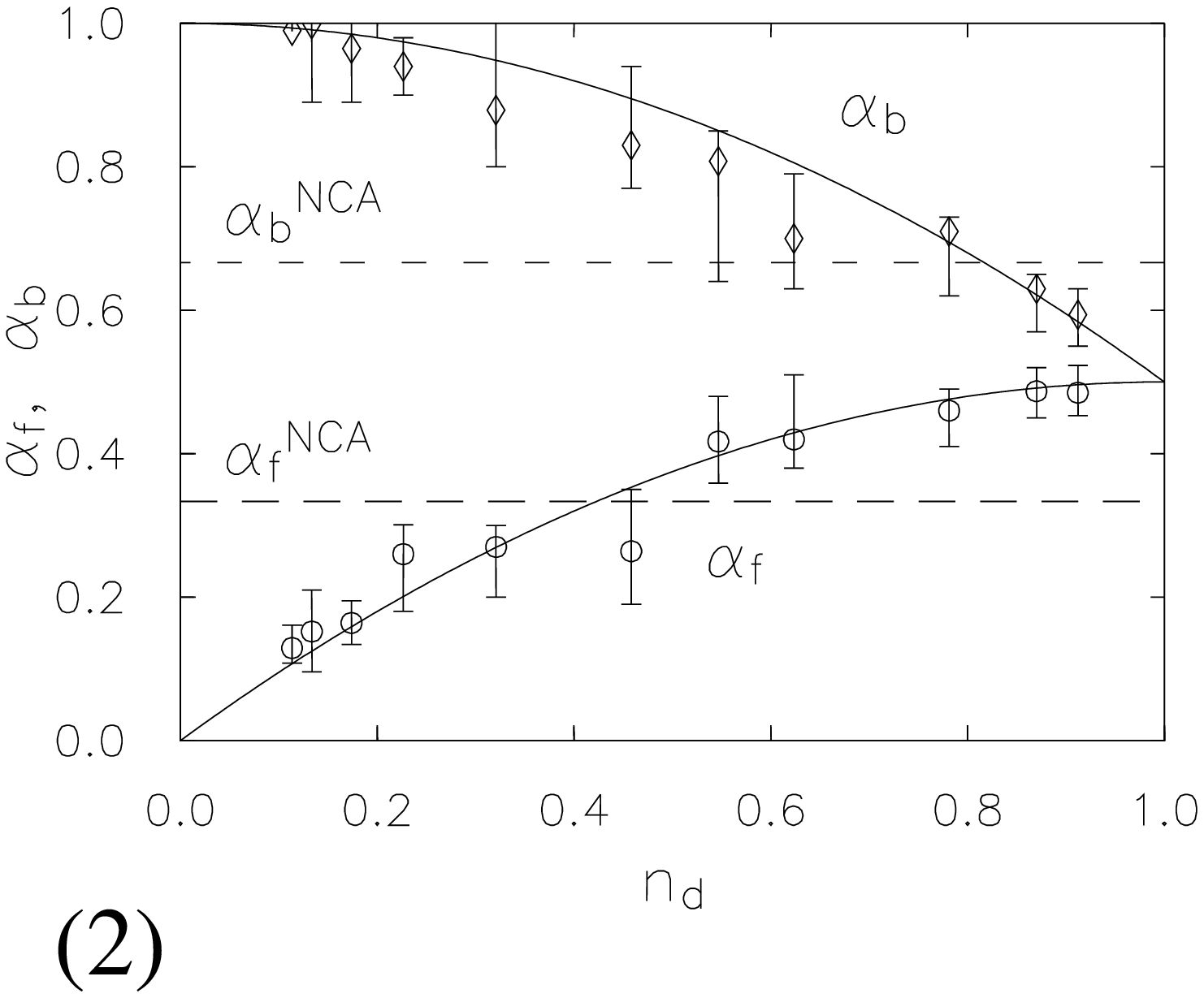,width=5.8cm}}
\vspace*{0.6cm}
\caption{ 
(1) Pseudofermion and slave boson spectral functions $A_f$ and $A_b$
in the Kondo regime ($N=2$; 
$E_d=-0.05$, $\Gamma =0.01$ in units of the half--bandwidth $D$), 
for a) the single--channel ($M=1$) and b) the
multi--channel ($M=2$) case.
In a) the symbols represent the results of NRG for the same 
parameter set, $T=0$. The slopes of the dashed lines indicate the exact
threshold exponents as derived in section 3 for $M=1$ and as given by 
conformal field theory for $M=2$. The insets show magnified power law 
regions. \ 
(2) CTMA results (symbols with error bars) for the threshold 
exponents $\alpha _f$ and $\alpha _b$ of $A_f$ and $A_b$, $N=2$, 
$M=1$. Solid lines: exact values (section 3), dashed lines: NCA results 
(section 4.2).
\label{spectralfb}}
\end{figure}

As shown in Fig.~\ref{spectralfb} (1) a), 
the auxiliary particle spectral functions obtained from CTMA 
\cite{kroha.97} are in 
good agreement with the results of a numerical renormalization 
group (NRG) calculation \cite{costi.94}
(zero temperature results), given the uncertainties in the 
NRG at higher frequencies. Typical behavior in the Kondo regime 
is obtained: a broadened peak in $A_b$ at $\omega\simeq |E_d|$, 
representing the hybridizing $d$--level and a structure 
in $A_f$ at $\omega \simeq T_K$. Both functions display power law 
behavior at frequencies
below $T_K$, which at finite $T$ is cut off at the scale $\omega
\simeq T$. The exponents extracted from the frequency range
$T<\omega<T_K$ of our finite $T$ results  
compare well with the exact result also shown
(see insets of Fig.~\ref{spectralfb} ($1 a$)). A similar analysis has been 
performed for a number of parameter sets spanning the complete range of
$d$--level occupation numbers $n_d$. The extracted power law exponents
are shown in Fig.~\ref{spectralfb} (2), 
together with error bars estimated 
from the finite frequency ranges over which the fit was made.
The comparatively large error bars
in the mixed valence regime arise because here spin flip and
charge fluctuation processes, described by the poles in $T^{(cf)}$ and
$T^{(cb)}$, respectively, are of equal importance, impeding the 
convergence of the numerical procedure. In this light, the agreement
with the exact results (solid curves) is very good, 
the exact value lying within the error bars or very close 
in each case. 

\begin{figure}
\vspace*{-0cm}
\centerline{\psfig{figure=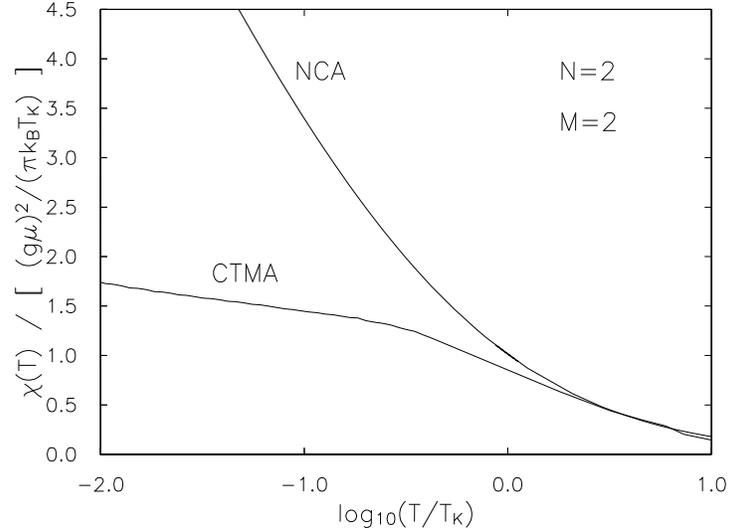,width=9.5cm}}
\vspace*{0.6cm}
\caption{
Static susceptibility of the two--channel Anderson impurity model:
CTMA and NCA results ($E_d=-0.8D$, $\Gamma = 0.1D$, Land\'e factor $g=2$).
\label{suscM=2}}
\end{figure}
In the multi--channel case ($N\geq 2$, $M\geq N$) NCA has been
shown \cite{coxruck.93} to reproduce asymptotically the correct 
threshold exponents, $\alpha _f = M/(M+N)$, $\alpha _b = N/(M+N)$,
in the Kondo limit. Calculating the T--matrices using NCA Green's 
functions (as discussed in the single--channel case) we find again
a pole in the singlet channel of $T^{(cf)}$. However, in this case 
the CTMA does not renormalize the NCA exponents 
in the Kondo limit of the two--channel model, i.e.~the threshold 
exponents obtained from the CTMA solutions are very close to the
exact ones, $\alpha _f =1/2$, $\alpha _b=1/2$, as shown in 
Fig.~\ref{spectralfb} (1) b). 
  
The agreement of the CTMA exponents with their exact values
in the Kondo, mixed valence and empty impurity regimes of the
single--channel model and in the Kondo regime of the
two--channel model  may be taken as
evidence that the T--matrix approximation correctly 
describes both the FL and the non--FL
regimes of the SU(N)$\times$SU(M) Anderson model (N=2, M=1,2).
Therefore, we expect the CTMA to correctly describe  
physically observable quantities of the SU(N)$\times$SU(M) Anderson 
impurity model as well. As a check, we have calculated
the static spin susceptibility $\chi$ of the two--channel Anderson 
model in the Kondo regime by solving the CTMA equations (see appendix B)
in a finite magnetic field $H$ coupled to the impurity spin and taking 
the derivative of the magnetization 
$M = \frac{1}{2}g \mu _B \langle n_{f\uparrow} - n_{f\downarrow}\rangle$ 
with respect to $H$. 
The resulting $\chi (T) = (\partial M / \partial H )_T$ is shown 
in Fig.~\ref{suscM=2}. It is seen that CTMA correctly reproduces
the exact \cite{wiegmann.83} logarithmic temperature dependence below 
the Kondo scale $T_K$. In contrast, the NCA solution recovers the 
logarithmic behavior only far below $T_K$. Other physical quantities 
will be calculated for the Anderson model in forthcoming work.

\section{Conclusion}
We have presented a novel technique
to describe correlated quantum impurity systems with strong onsite
repulsion, which is based on a conserving formulation of the 
auxiliary boson method. The conserving scheme allows to preserve the
conservation of the local charge $Q$ without taking into account 
time dependent fluctuations of the gauge field $\lambda$. 
Taking, as a result, $\lambda$ as a time independent quantity
represents a great simplification of this approach.
As a standard diagram technique this method has the potential to be 
applicable to problems of correlated systems on a lattice as well,
while keeping the full dynamics of the pseudofermion and slave boson 
fields. By including the leading infrared singular contributions
(spin flip and charge fluctuation processes),
this technique allows to correctly describe 
physical quantities, like the magnetic susceptibility, both in the
Fermi and in the non--Fermi liquid regime,
over the complete temperature range, including the crossover to the 
correlated many--body state at the lowest temperatures.  

We wish to thank S.~B\"ocker, 
T.A.~Costi, S.~Kirchner, A.~Rosch, A.~Ruckenstein and
Th.~Schauerte for stimulating discussions.
S.~B\"ocker has performed part of the numerical solutions. 
This work is supported by DFG through SFB 195 and by the
Hochlei\-stungsrechenzentrum J\"ulich through a grant of computer 
time on a Cray T3E parallel computer.

\vspace*{1.4cm}
\centerline{\bf Appendix}

\begin{appendix}

\section{Infrared cancellation of non--CTMA diagrams} 
The CTMA is not only justified on physical grounds by the inclusion 
of the maximum number of spin flip and charge fluctuation processes 
at any given order of perturbation theory, but also by an infrared
cancellation of all diagrams not included in the CTMA. In the following
we will prove this cancellation theorem.

\begin{figure}
\vspace*{-0cm}
\centerline{\psfig{figure=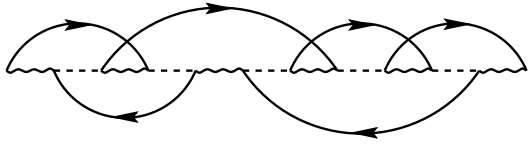,width=7cm}}
\vspace*{0.6cm}
\caption{
Typical pseudofermion self--energy skeleton diagram of loop order
$L=6$, containing $L_{sp}=1$ spin (or fermion) loop and $L_{ch}=1$
channel loop. 
\label{selfenergy}}
\end{figure}
(1) {\em Power counting}.---
Each auxiliary particle loop carries a factor of the fugacity 
${\rm exp}(-\beta\lambda)$, which vanishes upon projection onto the
$Q=1$ subspace, $\lambda\rightarrow \infty$. Therefore, an arbitrary 
$f$ or $b$ self--energy diagram consists of one single line of 
alternating fermion and boson propagators, with the hybridization 
vertices connected by conduction electron lines in any possible way, 
as shown in Fig.~\ref{selfenergy} (see also \cite{keiter.71}). 
Such a fermion self--energy skeleton diagram of loop order $L$ is 
calculated as
\begin{eqnarray}
\Sigma _{f}^{(L)}(\omega ) &=& (-
1)^{3L-1+L_{sp}}\;N^{L_{sp}}\;M^{L_{ch}}\;
\Gamma ^L\nonumber\\
&&\hspace*{-1.2cm}
\times\; 
\int\frac{ d\varepsilon _1}{\pi}\dots \frac{ d\varepsilon _L}{\pi}
\;f(\varepsilon _1) \dots \;f(\varepsilon _L) \
A^0_{c}(s_1\varepsilon _1)\dots A^0_{c}(s_L\varepsilon _L)
\label{eq:selfenergy}\\
&&\hspace*{-1.2cm}
\times\;G_b(\omega+\omega _1)G_f(\omega+\omega_1+\omega_1')\;\dots
\nonumber\\
&&\hspace*{-1.2cm}
\times\;\dots\;G_b\Big(\omega+\sum _{i=1}^{k}\omega _i+
                      \sum _{i=1}^{k-1}\omega _i'\Big)\;
       G_f\Big(\omega+\sum _{i=1}^{k}(\omega_i+\omega_i')\Big)\;\dots\;
       G_b(\omega+\omega _L)\; ,\nonumber
\end{eqnarray}
where $G_{f,b}$ are the {\em renormalized}\/, i.e. power law divergent 
auxiliary particle propagators, and
$L_{sp}$ and $L_{ch}$ denote the number of spin (or fermion, $c$--$f$) 
loops and the number of channel (or $c$--$b$) loops contained in the
diagram, respectively. Spin and channel indices are not shown
for simplicity. Each of the auxiliary particle 
frequencies $\omega _i$, $\omega _i'$ coincides
with one of the integration variables $\varepsilon _j$, $j=1,\dots , L$, 
in such a way that energy is conserved at each hybridization vertex.
This implies that the sign of the frequency carried by
a $c$--electron line is $s_i = +$, if the $c$--electron line runs from
right to left, and $s_i = - $, if it runs from left to right in 
Fig.~\ref{selfenergy}. An analogous expression holds for the
slave boson self--energy diagrams. By substituting 
$x_j = \varepsilon _j/\omega$, $j=1,\dots , L$ and factoring 
out $\omega ^{-\alpha_{f,(b)}}$ from each fermion (boson) propagator, 
the infrared behavior of the term Eq.~(\ref{eq:selfenergy}) 
is deduced as
\begin{eqnarray}
{\rm Im}
\Sigma _{f,b}^{(L)}(\omega ) = 
C \omega ^{\alpha _{f,b}+L(1-\alpha _f-\alpha _b)},
\label{eq:powercount}
\end{eqnarray}
where $C$ is a finite constant. Clearly, when the NCA solutions 
are inserted for the propagators $G_{f,b}$, 
i.e. $\alpha _f +\alpha _b = 1$, their power law behavior is just 
reproduced by any term of the form Eq.~(\ref{eq:selfenergy}). 
However, this is no longer the case for the exact propagators 
in the Fermi liquid regime ($M < N$), where in 
general $\alpha _f +\alpha _b > 1$.  
Thus, the infinite resummation of terms to arbitrary loop order is
unavoidable in this case.

\begin{figure}
\vspace*{-0cm}
\centerline{\psfig{figure=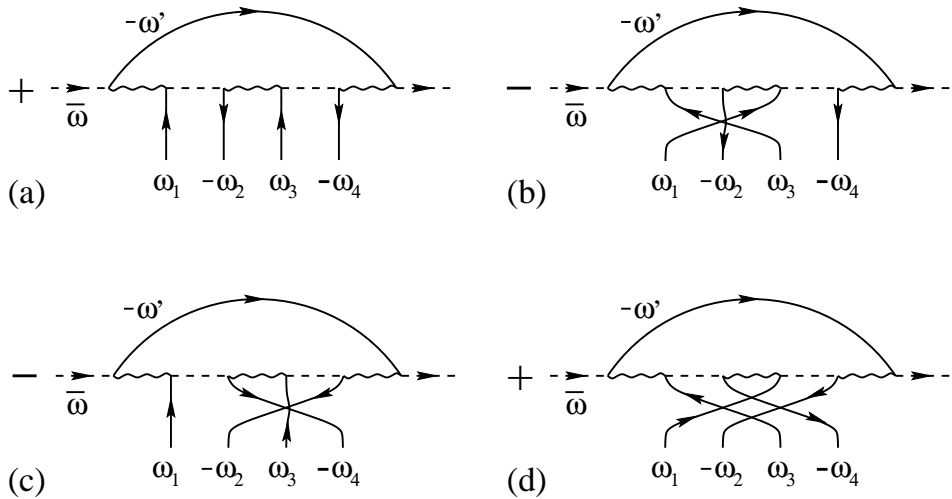,width=\textwidth}}
\vspace*{0.6cm}
\caption{
Set of contributions to skeleton diagrams {\em not}\/
contained in CTMA which cancel in the infrared limit
to leading and subleading order in the external frequency, 
$\omega \rightarrow 0$.   
\label{cancellation}}
\end{figure}
(2) {\em Infrared cancellation}.--- As discussed in section 5.2, the CTMA
is equivalent to the self--consistent summation of all skeleton free
energy diagrams, where a conduction electron line spans at most two
hybridization vertices (Fig.~\ref{CTMA}). 
Thus, any skeleton self--energy diagram {\em not}\/
included in CTMA contains at least one conduction electron ``arch''
which spans four (or more) vertices, with four conduction lines reaching
from inside to outside of the arch as shown in Fig.~\ref{cancellation}(a).
For each such diagram there exists another skeleton, which differs 
from Fig.~\ref{cancellation}(a) only in that the end points of two 
conduction lines inside the arch are interchanged 
(Fig.~\ref{cancellation}(b)). The corresponding permutation of
fermionic operators implies a relative sign between the terms
Fig.~\ref{cancellation}(a) and (b).
Without loss of generality we now assume
$\omega >0$ for the external frequency of the self--energy.
The leading infrared singular behavior of the term (\ref{eq:selfenergy})
arises from those parts of the integrations, where the arguments of
the $G_f$, $G_b$ are such that the divergences of all propagators lie
within the integration range. This implies at least 
$-\omega \leq \varepsilon _j \leq 0$, $j=1,\dots , L$. Therefore,
the terms corresponding to Fig.~\ref{cancellation}(a), (b)
differ only in the frequency arguments of the Green's functions inside
the arch, and at temperature $T=0$ the leading infrared behavior of their 
sum reads,
\begin{eqnarray}
\Sigma _{f}^{(L,a)}(\omega )+\Sigma _{f}^{(L,b)}(\omega ) 
&{\buildrel{_{\omega \to 0}}\over{=}}& 
(-1)^{3L-1+L_{sp}}\;N^{L_{sp}}M^{L_{ch}}
\Gamma ^L \times\\
&&\hspace*{-3.8cm}
\int _{-\omega}^0 \frac{ d\varepsilon _1}{\pi}
\dots\frac{ d\varepsilon _L}{\pi}\;F(\omega,\{\varepsilon _j\})\;
\big[ G_f(\bar\omega+\omega '+\omega _1)
      G_b(\bar\omega+\omega '+\omega_1+\omega_2)
\nonumber\\
&&\hspace*{+0.56cm}
-G_f(\bar\omega+\omega '+\omega _3)
 G_b(\bar\omega+\omega '+\omega_3+\omega_2)
\big]\nonumber
\label{eq:cancel}
\end{eqnarray}
Here $\bar\omega$ denotes the sum of all frequencies $\omega$,
$\varepsilon _j$ entering the diagrammatic part, 
Fig.~\ref{cancellation}, from the left, and $F(\omega,\{\varepsilon _j\})$ 
consists of all terms which are not altered by interchanging the
$c$--electron lines. In the infrared limit, $\omega _1-\omega _3 \to 0$,
the term in square brackets may be written as
\begin{eqnarray}
\frac{d}{d\bar\omega}\big[G_f(\bar\omega+\omega ')
                          G_b(\bar\omega+\omega '+\omega_2)\big]
                     (\omega _1-\omega _3)
\end{eqnarray}
and upon performing the integrations 
over $\omega _1$, $\omega _3$ the difference $(\omega _1-\omega _3)$ leads
to an additional factor of $\omega $.  
A similar cancellation of the leading infrared singularity occurs
between the terms shown in Fig.~\ref{cancellation}(c), (d).
In an analogous way it may be shown that combining the 
terms Fig.~\ref{cancellation} (a)--(d) leads to a factor of $\omega^2$
compared to the power counting result for one single term.
Thus, the infrared singularity of all non--CTMA terms of loop order $L$
is weaker than the $L$th order CTMA terms by at least $O(\omega ^2)$,
\begin{equation}
\Sigma _{f,b}^{(L,a)}(\omega )+\dots +\Sigma _{f}^{(L,d)}(\omega ) 
\ {\buildrel{_{\omega \to 0}}\over{\propto}}\ 
\omega ^{\alpha _{f,b}+L(1-\alpha _f-\alpha _b)+2}.
\end{equation}
It should be emphasized that in the above derivation, $L$ appears only
as a parameter and, thus, the cancellation theorem holds for 
arbitrarily high loop order $L$. This proves that the CTMA captures the 
leading and subleading infrared singularities ($\omega \to 0$)
at any given order $L$.

\section{CTMA equations}
In this appendix we give explicitly the self--consistent equations 
which determine the auxiliary particle self--energies within CTMA. 
For that purpose, it is useful to define conduction electron--fermion 
and conduction electron--boson vertex functions  
$T^{(cf)\,(\pm)}$, $T^{(cb)\,(\pm)}$ without ($+$) or 
with ($-$) an alternating sign between terms with even and odd number 
of rungs (compare Fig. \ref{tmat}). In the Matsubara representation,
these vertex functions, to be labelled ``even'' ($+$) and ``odd'' ($-$) 
below, are given by the following Bethe--Salpeter equations:
\begin{eqlettarray}
T^{(cf)\,(\pm)\,\mu}_{\phantom{(f)\,(\pm)\,}\sigma ,\tau}
(i\omega _n, i\omega _n ', i\Omega _n) &=&
U^{(cf)\,\mu}_{\phantom{(f)\,}\sigma ,\tau}
(i\omega _n, i\omega _n ', i\Omega _n) 
\nonumber\\
&&\hspace*{-1.6cm}
\pm V^2\frac{1}{\beta}\sum _{\omega _n''}
G_{b\bar\mu}(i\omega _n + i\omega _n '' - i\Omega _n ) \times 
\eqalabel{cftmatpm}\\
&&\hspace*{-1.6cm}
G_{f\sigma}(i\omega _n'') \ G^0_{c\mu\tau}(i\Omega _n -i\omega _n '')\ 
T^{(cf)\,(\pm)\,\mu}_{\phantom{(f)\,(\pm)\,}\tau ,\sigma}
(i\omega _n '', i\omega _n ', i\Omega _n ) \nonumber\\
U^{(cf)\,\mu}_{\phantom{(f)\,}\sigma ,\tau}
(i\omega _n, i\omega _n ', i\Omega _n) &=&
-V^4\frac{1}{\beta}\sum _{\omega _n''}
G_{b\bar\mu}(i\omega _n + i\omega _n '' - i\Omega _n ) \times 
\label{ucf}\\
&&\hspace*{-1.6cm}
G_{f\sigma}(i\omega _n'') \ G^0_{c\mu\tau}(i\Omega _n -i\omega _n '')\ 
G_{b\bar\mu}(i\omega _n ' + i\omega _n '' - i\Omega _n )\nonumber
\end{eqlettarray}\noindent
and
\begin{eqlettarray}
T^{(cb)\,(\pm)\sigma}_{\phantom{(b)(\pm)}\mu ,\nu}
(i\omega _n, i\omega _n ', i\Omega _n ) &=&
U^{(cb)\,\sigma}_{\phantom{(b)}\mu ,\nu}
(i\omega _n, i\omega _n ', i\Omega _n ) 
\nonumber\\
&&\hspace*{-1.6cm}
\pm V^2\frac{1}{\beta}\sum _{\omega _n''}
G_{f\sigma}(i\omega _n + i\omega _n '' - i\Omega _n ) \times 
\eqalabel{cbtmatpm}\\
&&\hspace*{-1.6cm}
G_{b\bar\mu}(i\omega _n'') \ G^0_{c\nu\sigma}(i\omega _n ''-i\Omega _n )\ 
T^{(cb)\,(\pm)\sigma}_{\phantom{(b)(\pm)}\nu ,\mu}
(i\omega _n '', i\omega _n ', i\Omega _n ) \nonumber\\
U^{(cb)\,\sigma}_{\phantom{(b)}\mu ,\nu}
(i\omega _n, i\omega _n ', i\Omega _n ) &=&
-V^4\frac{1}{\beta}\sum _{\omega _n''}
G_{f\sigma}(i\omega _n + i\omega _n '' - i\Omega _n ) \times 
\label{ucb}\\
&&\hspace*{-1.6cm}
G_{b\bar\mu}(i\omega _n'') \ G^0_{c\nu\sigma}(i\omega _n ''-i\Omega _n )\ 
G_{f\sigma}(i\omega '_n + i\omega _n '' - i\Omega _n ). \nonumber
\end{eqlettarray}\noindent
Note that, in addition to the alternating sign, 
these vertex functions differ from
the T--matrices defined in Eqs.~(\ref{cftmateq}), (\ref{cbtmateq})
in that they contain only terms with two or more rungs, since
the inhomogeneous parts $U^{(cf)}$ and $U^{(cb)}$ represent terms with
two bosonic or fermionic rungs, respectively. The terms with a 
single rung correspond to the NCA diagrams and are evaluated separately 
(see below).

The spin degrees of freedom of $T^{(cf)\,(\pm)}$ are uniquely 
determined by the spin indices $\sigma$, $\tau$ of the ingoing 
conduction electron and pseudofermion lines (Fig.~\ref{tmat}). 
It is instructive to note that in the spin $S=1/2$ case ($N=2$)
the singlet and triplet vertex functions  
(which correspond to the two--particle Green's functions in the singlet
channel, $\phi ^{\;s} \sim \sum _\sigma 
\langle T\{(c_{\sigma}f_{-\sigma}-c_{-\sigma}f_{\sigma})
           (c^{\dag}_{\sigma}f^{\dag}_{-\sigma}-
            c^{\dag}_{-\sigma}f^{\dag}_{\sigma})\}\rangle$,
and in the triplet channel with magnetic quantum number $m=0,\pm 1$,  
$\phi ^{\;t}_{m=0} \sim \sum _\sigma 
\langle T\{(c_{\sigma}f_{-\sigma}+c_{-\sigma}f_{\sigma})
        (c^{\dag}_{\sigma}f^{\dag}_{-\sigma}+
         c^{\dag}_{-\sigma}f^{\dag}_{\sigma})\}\rangle$,
$\phi ^{\;t}_{m=\pm 1} \sim  
\langle T\{c_{\pm\frac{1}{2}}f_{\pm\frac{1}{2}}
           c^{\dag}_{\pm\frac{1}{2}}f^{\dag}_{\pm\frac{1}{2}}\}\rangle$, 
respectively)
may be identified in the following way,
\begin{eqlettarray}
T^{(cf)\, s} &=&        \sum _{\sigma } T^{(cf)\, (-)}_{\sigma , -\sigma} 
\\
T^{(cf)\, t}_{m=0} &=&    \sum _{\sigma } T^{(cf)\, (+)}_{\sigma , -
\sigma} \\
T^{(cf)\, t}_{m=\pm 1}&=& \phantom{\sum _{\sigma }} 
                        T^{(cf)\, (+)}_{\pm\frac{1}{2},\pm\frac{1}{2} }. 
\end{eqlettarray}
\begin{figure}
\vspace*{-0cm}
\centerline{\psfig{figure=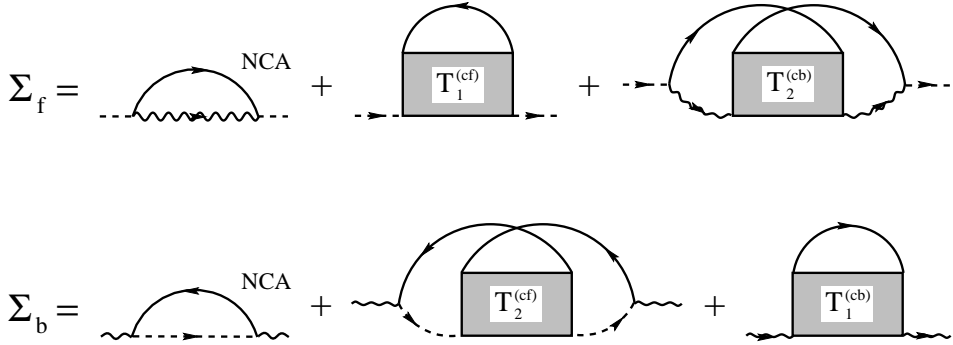,width=\textwidth}}
\vspace*{0.6cm}
\caption{Diagrammatic representation of the CTMA expressions for
pseudoparticle self--energies $\Sigma _f$ and $\Sigma _b$.
The first term drawn on the righthand side
of $\Sigma _f$ and $\Sigma _b$, respectively, is the NCA diagram. 
The diagrammatic parts $T^{(cf)}_{1,2}$, $T^{(cb)}_{1,2}$ are 
explained in the text.
\label{selfenergyCTMA}}
\end{figure}
Analogous relations hold for the conduction electron--boson
vertex function in terms of the channel degrees of freedom $\mu$, $\nu$.
The total CTMA pseudoparticle self--energies, as derived by functional 
differentiation from the generating functional $\Phi$, Fig.~\ref{CTMA},
are shown in Fig.~\ref{selfenergyCTMA} and consist of three terms each,
\begin{eqnarray}
\Sigma _{f\sigma}(i\omega  _n) &=&\Sigma _{f\sigma}^{(NCA)}(i\omega _n )+
                             \Sigma _{f\sigma}^{(cf)} (i\omega _n ) +
                             \Sigma _{f\sigma}^{(cb)} (i\omega _n )\\
\Sigma _{b\bar\mu}(i\omega _n)&=&\Sigma _{b\bar\mu}^{(NCA)}(i\omega _n )+
                             \Sigma _{b\bar\mu}^{(cf)} (i\omega _n ) +
                             \Sigma _{b\bar\mu}^{(cb)} (i\omega _n )\; .
\end{eqnarray}
The first term of $\Sigma _f$ and $\Sigma _b$ represents  
the NCA self--energies, Eqs.~(\ref{sigfNCA}), 
(\ref{sigbNCA}). The second and third terms arise from the spin and 
the charge fluctuations, respectively, and are given for pseudofermions by
\begin{eqlettarray}
\Sigma _{f\sigma}^{(cf)}(i\omega _n) &=& M\;\frac{1}{\beta}\sum _{\Omega _n}
G^0_{c}(i\Omega _n-i\omega _n)
T^{(cf)}_1 (i\omega _n,i\omega _n,i\Omega _n)\\
\Sigma _{f\sigma}^{(cb)}(i\omega _n) &=& 
-M\; V^2\frac{1}{\beta ^2}\sum _{\omega _n '\omega _n''}
G^0_{c}(i\omega _n-i\omega _n')
G _{b}(i\omega _n')\times\\
&&
T^{(cb)}_2 
(i\omega _n ',i\omega _n '',i\omega _n'+i\omega _n''-i\omega _n)
G^0_{c}(i\omega _n-i\omega _n'') G _{b}(i\omega _n'')\nonumber
\end{eqlettarray}
and for slave bosons by
\begin{eqlettarray}
\Sigma _{b\bar\mu}^{(cf)}(i\omega _n ) &=& 
-N\; V^2\frac{1}{\beta ^2}\sum _{\omega _n '\omega _n''}
G^0_{c}(i\omega _n'-i\omega _n)
G _{f}(i\omega _n')\times\\
&&
T^{(cf)}_2 (i\omega _n ',i\omega _n '',i\omega _n'+i\omega _n''-i\omega _n)
G^0_{c}(i\omega _n''-i\omega _n)
G _{f}(i\omega _n'')\nonumber\\
\Sigma _{b\bar\mu}^{(cb)}(i\omega _n) &=& N\;\frac{1}{\beta}\sum _{\Omega _n}
G^0_{c}(i\omega _n -i\Omega _n) T^{(cb)}_1 
(i\omega _n,i\omega _n,i\Omega _n)\; ,
\end{eqlettarray}
where the vertex functions appearing in these expressions are defined as
\begin{eqlettarray}
T^{(cf)}_1 &=& \frac{N+1}{2}\;T^{(cf)\,(+)} + \frac{N-1}{2}\; T^{(cf)\,(-)}
               -N\, U^{(cf)}\label{tcf1}\\
T^{(cf)}_2 &=& \frac{N+1}{2}\;T^{(cf)\,(+)} - \frac{N-1}{2}\; T^{(cf)\,(-)}
               - U^{(cf)} \label{tcf2}
\eqalabel{tcf12}
\end{eqlettarray}
\begin{eqlettarray}
T^{(cb)}_1 &=& \frac{M-1}{2}\; T^{(cb)\,(+)} + \frac{M+1}{2}\; 
               T^{(cb)\,(-)}-M\, U^{(cb)}\label{tcb1}\\
T^{(cb)}_2 &=& \frac{M-1}{2}\; T^{(cb)\,(+)} - \frac{M+1}{2}\; 
               T^{(cb)\,(-)}-U^{(cb)}\; .\label{tcb2}
\eqalabel{tcb12}
\end{eqlettarray}
These combinations of the even and odd vertex
functions ensure the proper spin and channel summations in the
self--energies. For the sake of clarity, the spin and channel 
indices as well as the frequency variables are not shown explicitly. 
In Eqs. (\ref{tcf1}), (\ref{tcb1}) 
the terms with two rungs, $N\, U^{(cf)}$, $M\, U^{(cb)}$, have been
subtracted, since they would generate non--skeleton self--energy diagrams.
Likewise, in Eqs. (\ref{tcf2}), (\ref{tcb2}) the two--rung terms have
been subtracted in order to avoid a double counting of terms in the
self--energies. 

We now turn to the analytic continuation to real frequencies of the
expressions derived above. 
Transforming the Matsubara summations into contour integrals shows
that integrations along branch cuts of
auxiliary particle Green's functions carry an additional factor
${\rm exp}(-\beta\lambda)$ as compared to integrations along branch 
cuts of physical Green's functions, which vanishes upon projection 
onto the the physical Fock space, $\lambda \rightarrow \infty$.
Thus, as a general rule, only integrations along branch cuts of the
$c$--electron propagators contribute to the auxiliary particle 
self--energies. Therefore, by performing the analytic continuation, 
$i\omega_n \rightarrow \omega -i0 \equiv \omega$ in all frequency
variables, we obtain the advanced pseudofermion self--energy,
\begin{eqlettarray}
\Sigma _{f\sigma}^{(cf)}(\omega ) &=& M
\int \frac{ d\varepsilon}{\pi}\;f(\varepsilon -\omega)\;
A^0_{c}(\varepsilon -\omega)\; \pi{\cal N}(0)
T^{(cf)}_1 (\omega ,\omega ,\varepsilon )\\
\Sigma _{f\sigma}^{(cb)}(\omega ) &=& 
-M\; \Gamma \int \frac{ d\varepsilon}{\pi}\int \frac{ d\varepsilon '}{\pi}
\;f(\varepsilon -\omega)\;f(\varepsilon ' -\omega)\;\times\\
&&\hspace*{-1.2cm}
A^0_{c}(\omega -\varepsilon )
G _{b}(\varepsilon )\; 
\pi{\cal N}(0) T^{(cb)}_2 (\varepsilon ,\varepsilon ' ,
\varepsilon +\varepsilon ' -\omega )
A^0_{c}(\omega -\varepsilon ') G _{b}(\varepsilon ' )\nonumber
\eqalabel{sigfcontinued}
\end{eqlettarray}
and the advanced slave boson self--energy,
\begin{eqlettarray}
\Sigma _{b\bar\mu}^{(cf)}(\omega ) &=& 
-N\; \Gamma \int \frac{ d\varepsilon}{\pi}\int \frac{ d\varepsilon '}{\pi}
\;f(\varepsilon -\omega)\;f(\varepsilon '-\omega )\;\times\\
&&\hspace*{-1.2cm}
A^0_{c}(\varepsilon -\omega )
G _{f}(\varepsilon )\;
\pi{\cal N}(0)T^{(cf)}_2 (\varepsilon ,\varepsilon ' ,
\varepsilon +\varepsilon ' -\omega )
A^0_{c}(\varepsilon ' -\omega)
G _{f}(\varepsilon ' )\nonumber\\
\Sigma _{b\bar\mu}^{(cb)}(\omega ) &=& -N
\int \frac{ d\varepsilon}{\pi}\;f(\varepsilon -\omega)\;
A^0_{c}(\omega -\varepsilon )\;
\pi{\cal N}(0) T^{(cb)}_1 (\omega ,\omega ,\varepsilon ) ,
\eqalabel{sigbcontinued}
\end{eqlettarray}
where the vertex functions are given by Eqs. (\ref{tcf12}), 
(\ref{tcb12}) with
\begin{eqlettarray}
T^{(cf)\,(\pm)\,\mu}_{\phantom{(f)\,(\pm)\,}\sigma ,\tau}
(\omega , \omega ' , \Omega ) &=&
U^{(cf)\,(\pm)\,\mu}_{\phantom{(f)\,(\pm)\,}\sigma ,\tau}
(\omega , \omega ' , \Omega )
\nonumber\\
&&\hspace*{-2.0cm}
\pm (-\Gamma) \int \frac{ d\varepsilon}{\pi}\;f(\varepsilon -
\Omega)\;\times \\
&&\hspace*{-2.0cm}
G_{b\bar\mu}(\omega+\varepsilon-\Omega)  
G_{f\sigma}(\varepsilon )  A^0_{c\mu\tau}(\Omega -\varepsilon)  
T^{(cf)\,(\pm)\,\mu}_{\phantom{(f)\,(\pm)\,}\tau ,\sigma}
(\varepsilon , \omega ', \Omega  ) \nonumber\\
\pi{\cal N}(0)\; U^{(cf)\,(\pm)\,\mu}_{\phantom{(f)\,(\pm)\,}\sigma ,\tau}
(\omega , \omega ' , \Omega ) &=&
+\Gamma ^2 \int \frac{ d\varepsilon}{\pi}\;f(\varepsilon -\Omega) \times\\
&&\hspace*{-2.0cm}
G_{b\bar\mu}(\omega +\varepsilon -\Omega) 
G_{f\sigma}(\varepsilon )  A^0_{c\mu\tau}(\Omega -\varepsilon) 
G_{b\bar\mu}(\omega '  +\varepsilon -\Omega)  \nonumber
\eqalabel{tcfcontinued}
\end{eqlettarray}
and
\begin{eqlettarray}
T^{(cb)\,(\pm)\sigma}_{\phantom{(b)(\pm)}\mu ,\nu}
(\omega , \omega ', \Omega ) &=&
U^{(cb)\,(\pm)\sigma}_{\phantom{(b)(\pm)}\mu ,\nu}
(\omega , \omega ', \Omega )
\nonumber\\
&&\hspace*{-2.0cm}
\pm (+\Gamma) \int \frac{ d\varepsilon}{\pi}\;f(\varepsilon -\Omega )\;
\times\\
&&\hspace*{-2.0cm}
G_{f\sigma}(\omega +\varepsilon -\Omega ) 
G_{b\bar\mu} (\varepsilon )
A^0_{c\nu\sigma}(\varepsilon -\Omega )\ 
T^{(cb)\,(\pm)\sigma}_{\phantom{(b)(\pm)}\nu ,\mu}
(\varepsilon , \omega ', \Omega ) \nonumber\\
\pi{\cal N}(0)\; U^{(cb)\,(\pm)\sigma}_{\phantom{(b)(\pm)}\mu ,\nu}
(\omega , \omega ', \Omega ) &=&
-\Gamma ^2\int \frac{ d\varepsilon}{\pi}\;f(\varepsilon -\Omega) \times\\
&&\hspace*{-2.0cm}
G_{f\sigma}(\omega + \varepsilon -\Omega ) 
G_{b\bar\mu}(\varepsilon ) A^0_{c\nu\sigma}(\varepsilon -\Omega)\ 
G_{f\sigma}(\omega '+ \varepsilon -\Omega ) \nonumber
\eqalabel{tcbcontinued}
\end{eqlettarray}\noindent
In the above expressions, like in the NCA equations (\ref{eq:NCA}),
we have used the dimensionless conduction electron spectral density, 
$A_{c}^0(\omega )=\frac{1}{\pi}\, {\rm Im}G_{c\mu\sigma}^0(\omega -i0)/
{\cal N}(0)$, and we have suppressed obvious spin and channel indeces. 
All frequency variables are to be understood as the limit
$\omega\equiv \omega -i0$. \\
The equations (\ref{sigfcontinued}), (\ref{sigbcontinued}), 
supplemented by the vertex functions Eqs. (\ref{tcf12}), (\ref{tcb12}), 
(\ref{tcfcontinued}), (\ref{tcbcontinued}) form, together with the 
NCA contributions Eqs. (\ref{sigfNCA}), (\ref{sigbNCA}) and the 
definitions of the auxiliary particle Green's functions, 
Eqs.~(\ref{green}), (\ref{green0}), the closed set of self--consistent
CTMA equations \cite{kroha.97}. 
It is seen that in these equations only those branches of the 
T--matrix vertex functions appear which are advanced with respect to all 
three frequency variables, although in general the T--matrix consists of
$2^3$ independent analytical branches. This simplification is a 
consequence of the exact projection onto the physical sector of Fock space. 
Inspection of the analytically continued CTMA equations also shows that 
the slave boson self--energy is obtained from the pseudofermion 
self--energy, including the proper signs, by simply replacing 
$G_f \leftrightarrow G_b$ and inverting the frequency argument of 
$A_c^0$ in all expressions.

Eqs. (\ref{sigfcontinued})--(\ref{tcbcontinued}) 
may be rewritten in terms of the spectral functions without
threshold, $\tilde A_{f,b}$, in a straight--forward way 
as explained in section 4.3, thus avoiding divergent statistical 
factors in the $d$--electron Green's function. The CTMA equations
are solved numerically by iteration. In each iteration step the
NCA and the T--matrix contributions are computed separetely. We solve the
linear T--matrix Bethe--Salpeter--Equations (\ref{tcfcontinued}),  
(\ref{tcfcontinued}) by discretization of the frequency integrals and
subsequent solution of the resulting multidimensional linear equations.
Because of the formulation given above, 
where first the vertex parts $T^{(cf)\;(\pm)}$ and
$T^{(cf)\;(\pm)}$, comprised of all T--matrix diagrams with two or more rungs,
are calculated and in a second step the non--skeleton two--rung contributions
are subtracted (Eqs.~(\ref{tcf12}), (\ref{tcb12})), each term of the 
T--matrix equations involves at most one frequency integration.
The resulting numerical effort is manageable: One complete 
iteration within the self--consistent scheme has been done on a parallel
computer within approximately 5--10 s CPU time.

\end{appendix}

\end{document}